\documentclass[prb,twocolumn]{revtex4-1} 

\usepackage{graphicx}
\usepackage{amsmath}
\usepackage{amsfonts} 
\usepackage{mathrsfs} 
\usepackage{amssymb}
\usepackage{color}
\usepackage{bm}
\usepackage{comment}

\def\beq{\begin{equation}}
\def\eeq{\end{equation}}
\def\barray{\begin{eqnarray}}
\def\earray{\end{eqnarray}}
\def\<{\langle}
\def\>{\rangle}

\def\myi{\imath}
\usepackage[margin=0.8in]{geometry}

\def\nq{N_{\rm 4}}
\newcommand{\numero}[1]{{\it (#1)}}

\newcommand{\order}{{\cal O}}

\newcommand{\di}{\mathrm{d}}

\begin{document}

\title{Statistical physics of nonlinear wave interaction}

\author{F. Antenucci$^{1,2}$, M. Ib\'a\~nez Berganza$^{1}$, L. Leuzzi$^{1,2}$} 

\affiliation{ $^1$IPCF-CNR, UOS Rome {\em Kerberos}, Piazzale Aldo
  Moro 5, I-00185, Roma, Italy\\ $^2$ Dipartimento di Fisica,
  Universit\`a di Roma ``Sapienza,''Piazzale Aldo Moro 5, I-00185,
  Roma, Italy }

\begin{abstract}
The thermodynamic  properties of vector ($O(2)$ and
Complex Spherical) models with four-body interactions are
analyzed. When defined in dense topologies, these are effective models
for the nonlinear interaction of scalar fields in the presence of a
stochastic noise, as has been well established for the case of the
mode locking laser formation in a closed cavity.
With the help of a novel efficient Monte Carlo algorithm we show how beyond the fully connected case novel and rich phenomenology emerges. 
Below a certain dilution threshold, the spherical model condensates in a non-equipartite way, while in the XY model the transition becomes continuous and the $O(2)$ symmetry remains unbroken, 
we attribute this fact to the invariance under local gauge transformations. The introduction of topological inhomogeneities in the network of quadruplets induces novel features: again symmetry conservation; 
the vanishing of two-point correlators; and a dynamical correlation function presenting two timescales, the large one being related to the transition between different degenerated configurations, 
connected by nonlocal gauge transformations. 
We discuss possible experimental implications of these results in the context of nonlinear optics.
\end{abstract}

\maketitle


\section{Introduction}
\label{sec:intro}

\subsection{Thermodynamic approach to nonlinear optics}

In the last decade there have been several fascinating attempts to understand nonlinear wave phenomena as collective, emergent
behavior.\cite{Fratalocchi2008,El2005,Rasmussen2000,Krumhansl1975,Gordon2002,Gordon2003,Vodonos2004,GordonFisherActiveML,GordonFisherOC,Katz2006}
Within such a scheme, the focus is not on the kinetics of the nonlinear wave propagation,\cite{Picozzi2014} but  on the
description in terms of static quantities in a suitably defined ensemble, in such a way that different wave regimes are in
correspondence with different thermodynamic phases of a Hamiltonian model.  

A set of fundamental works in this context are
Refs. \onlinecite{Gordon2002,Gordon2003,Vodonos2004} which describe
the mechanism of Passive Mode-Locking in multi-mode lasers within a
statistical mechanical framework.  The electromagnetic modes in this
case are the longitudinal modes of the resonant cavity, and the
non-linearity is provided by a {\it saturable absorber}, a device
which enhances high electromagnetic field intensity, hence favoring
modes with large amplitude and locked phases. The temporal evolution
of modes is described by a master equation \cite{HausPaper} accounting
for the nonlinear coupling of {\it tetrads} of modes (a four body
interaction), and with an additional stochastic drift term due to the
spontaneous emission, which opposes mode-locking as it tends to
incoherently disorder moduli and phases. In the limit in which the
dispersion can be neglected,\cite{GordonFisherOC} the master equation
leads to a Hamiltonian formulation such that the electromagnetic modes
can be regarded as (complex) spin degrees of freedom, coupled by a
four-body ferromagnetic interaction, while the stability of the system
is assured by a global constraint on the sum of the mode intensities
(a spherical constraint, in the spin language).  The steady state of
the laser is described by measurements in the canonical ensemble of
the spin model, where the role of the temperature is played by the
inverse square of {\it pumping rate} of the laser source.  The methods
of statistical physics applied to this problem reveal that, for
sufficiently high ratio between the pumping rate and the noise strength, a discontinuous
transition separating a para- from a ferromagnetic phase takes
place.\cite{PhysRevLett.73.3395} In the ferromagnetic regime the
phases and intensities of modes at different frequencies become {\it
  locked}, i.e., correlated, and long-range order appears, associated
with $O(2)$ symmetry breaking. In the optical language this phase
corresponds to a coherent light regime in which ultra-short
electromagnetic pulses are generated (the {\it Mode Locked} (ML) {\it
  regime}). On the other hand, if the spontaneous emission dominates,
light is in an {\it Incoherent Wave} (IW) regime with low power efficiency and flat intensity
spectrum, which is described by
a paramagnetic state in the spin language. This approach allows for a
treatment of the non-perturbative influence of noise, and explains the
discontinuous nature of the mode-locking transition, along with other
properties reminiscent of discontinuous transitions, as an hysteresis
effect called \emph{optical bistability}.\cite{Boyd2008} Variations of this
problem have also been considered, as the Active Mode
Locking,\cite{GordonFisherActiveML} injection of pulses from an
external source,\cite{PhysRevLett.95.013903} and a general agreement
with experimental results has been found.

On the other hand, there have been a series of theoretical works
generalizing the study of these Hamiltonians through the addition of
quenched disorder in the interaction
couplings.\cite{Angelani2006,Leuzzi2009,Conti2011,Antenucci2014} 
These more complex models may represent different physical situations, as
the random laser phenomena,\cite{wiersma2008physics} under specific
assumptions.\cite{AntenucciThesis} 
In this case, a sufficiently large amount of
disorder eventually leads to a glassy phase in the spin model, anticipated by a region
with nonzero complexity, which is believed to describe a frustrated
laser regime with absence of long-range correlations, possibly present in random lasers.

In the relevant statistical models in this context, the
electromagnetic modes are complex degrees of freedom (or $O(2)$ spins,
if their amplitude dynamics can be ignored) subject to a 4-body
interaction which can be purely ferromagnetic or disordered. These
are, in substance, XY or Complex Spherical $p$-spin ferromagnets or
spin glasses, with $p=4$. 
They have been studied so far in the mean field approximation, 
which is basically exact in the the fully connected case.
In this work we perform a systematic study of the
thermodynamics of the XY and spherical models with four body
interactions beyond mean field, considering the
influence of dilute topologies and of network correlations. From the
optical point of view, such a generalization allows to account for two
ingredients of crucial importance in optical systems, that could not
have been considered in previous studies, in which correlations were
disregarded. 

First, the role of mode frequencies, $\{\omega_n\}$,
which have an essential influence on the list of interacting mode tetrads  
since these are subject to an energy conservation prescription
on their four frequencies, called {\it frequency matching condition}
(see the next subsection). In cavity lasers it is not physically
justified to neglect the influence of mode frequencies. We will see
that they may induce correlations in the system dynamics which lead to
dramatic differences with respect to the mean field case, and these
differences have clear physical consequences in the optical counterpart.  

Second, the presence of dilution in the interaction network, 
from the fully connected down to the 
sparse network with an  extensive number of tetrads. 
This element is necessary, e.g., to account for the onset of lasing in more complicated
experimental setups as the random laser, in which the interaction
sparseness depends on the spatial superposition of the electromagnetic
fields of the modes. 
As we will see, a sufficiently large degree of dilution
induces nontrivial changes in the nature of the XY transition. Furthermore, in
the spherical model case, the dilution induces a transition to a regime
in which the nonlinearity prevents the equipartition of energy. Since the
seminal work of Fermi, Pasta and Ulam,\cite{Fermi1974} the
non-equipartition of energy induced by nonlinearity is one of the
crucial phenomena in nonlinear physics that claims for a statistical
treatment.\cite{Rasmussen2000,Fratalocchi2008}

From the point of view of statistical mechanics, on the other hand,
the models investigated in this article are novel, and present a
surprisingly rich phenomenology when fluctuations are allowed to take
place. 
As we explain in subsection
\ref{sec:gauge}, the Ising model with $p=4$ has already been considered
beyond mean field approximation, exhibiting slow dynamics and other kinetic features characteristic of glass formers. 
On the other hand, the XY model with suitable plaquette interactions is an effective lattice model for the gauge $O(2)$ field theory 
describing electromagnetism, but, to our knowledge, this is the first work considering this model within a statistical mechanical framework.
We will show how in the arena of
these models one can find, according to the dilution and to the presence
or absence of topological correlations, a variety of phenomenology
ranging from the symmetry conservation (reminiscent of the
Kosterlitz-Thouless transition), to different orders of the transition,
non-equipartite energy localization, and slow dynamics, among other
features.

To better motivate the study of these models in the optical context,
we review in some detail the Hamiltonian approach to the Passive mode
locking transition in the next subsection. Subsection \ref{sec:gauge}
is to review precedent studies of four-body interactions in
statistical physics. We then define the models under study and
describe their properties in Sec. \ref{sec:models}. The effect of a
sufficiently large amount of dilution on them is described in
Sec. \ref{sec:threshold}. Sec. \ref{sec:numerical} is dedicated to the
numerical methods that we have employed, and the consequent results
about the spherical and XY models are exposed in
Secs. \ref{sec:resultsSM}, \ref{sec:resultsXY} respectively. We will,
then, draw some analogies between these results and similar phenomena
occurring in lattice gauge theories (Sec. \ref{sec:analogy}), and
propose possible physical consequences in the field of nonlinear
optics in Sec. \ref{sec:optics}. Our conclusions are in
Sec. \ref{sec:conclusions}.

\subsection{Statistical approach to Mode Locking}
\label{sec:modelocking}
The evolution of the electromagnetic mode $a_l\in \mathbb{C}$, in a
standard passive mode locking laser is expressed through the well-known
master equation \cite{HausPaper}
\begin{eqnarray}
\label{eq:master_equation_haus}
 \frac{\di}{\di t} a_l (t) =&& 
 \left( G_l + \myi D_l \right) a_l (t) + 
 \\ 
 & &+ \left( \Gamma - \myi \Delta \right) 
   {\sum_{ k_1, k_2 , k_3  }}'
     a_{k_1} (t)  
 a_{k_2}^* (t)  
 a_{k_3} (t) + F_l(t) \, ;
 \nonumber
\end{eqnarray}
here the parameter $G_l$ represents the difference between the gain
and loss of the mode $l$ in a complete round-trip through the cavity,
$D_l$ is the group velocity dispersion of the wave packet, $\Gamma$ is
the nonlinear self-amplitude modulation coefficient associated to a
saturable absorber and, hence, to the passive mode-locking, and
$\Delta$ is the self-phase modulation coefficient (responsible of the
Kerr lens effect).  The noise $F_l(t)$ is generally assumed Gaussian,
white and uncorrelated:
\begin{align}
\nonumber
 & \langle F_{k_1}^* (t_1) \, F_{k_2} (t_2) \rangle = 2 T_0 \, \delta_{k_1 k_2 } \, \delta (t_1-t_2) \, ,
 \\
 & \langle F_{k_1} (t_1) \, F_{k_2} (t_2) \rangle = 0 \, ,
 \label{eq:uncorrelated_noise}
\end{align}
where $T_0$ is the spectral power of the noise.

A fundamental element, that deserves a particular attention in this
paper, is that the sum in the nonlinear term in
Eq. (\ref{eq:master_equation_haus}) is restricted to the tetrads of
modes such that the following {\it Frequency Matching Condition} (FMC)
\begin{align}
| \omega_{l} - \omega_{k_2} + \omega_{k_{3}} - \omega_{k_{4}} | \lesssim \gamma 
\label{eq:FMC}
\end{align}
is satisfied, where $\gamma$ is the single mode line-width.

In the following we are interested in the purely dissipative case, in which
the group velocity dispersion and the Kerr effect can be neglected.  
This includes the important case of soliton lasers.\cite{GordonFisherOC} 
The purely dissipative situation plays an exceptional role in our approach:
in this case, the
evolution depicted by Eq. (\ref{eq:master_equation_haus}) is
Hamiltonian, while the system remains stable because the gain decreases as
the optical intensity increases.\cite{Chen_94} To study the
equilibrium properties of the model, this last element can be included considering
an equivalent variant of the model where the gain assumes
the value that exactly keeps the total {\it optical power}, ${\cal
  E}=\sum_j |a_j|^2$ constant of motion, as Gordon and Fischer have
proposed in Ref \onlinecite{Gordon2002}.  In this way the system
evolves over the hypersphere:
\beq
\sum_j |a_j|^2 \equiv \epsilon N
\label{sphericalconstraint}
.
\eeq
In this situation, the effective temperature in the statistical model
is inversely proportional to the squared optical power: $T\equiv
T_0/\epsilon^2 $, where $T_0$ is the true heat-bath temperature.
Equivalently, the parameter that drives the transition in the photonic
system can be expressed through the so-called {\it pumping rate}
$\mathcal{P}^2 = T^{-1}$.


\subsection{Previous studies of 4-body models: lattice gauge theories}
\label{sec:gauge}
Ising models with four body (lattice plaquette) interactions have been
studied as cut-off regularized versions of scalar gauge
theories.\cite{Kogut1979} If the interacting quadruplets are suitably
defined in terms of plaquettes of a hyper-cubic lattice in $d$
dimensions, the model energy becomes invariant under flipping the sets
of four neighboring spins (a local gauge transformation).\footnote{in
  particular, if the quadruplets are formed by the four spins living
  in the edges of single plaquettes of an auxiliary hyper-cubic
  lattice, the local gauge transformation consists in flipping all
  spins corresponding to the edges incoming a given node of the
  auxiliary lattice.} The $p=4$ Ising model so defined is called 
  \emph{Ising lattice gauge theory} and is known to present a single, disordered
phase for any nonzero temperature in $d=2$, when it is equivalent to
an independent set of $d=1$ pairwise Ising models. In $d=3$ the Ising
lattice gauge theory exhibits a phase transition, which is related to
the $d=3$ Ising model transition.
The low-temperature phase is, however, unmagnetized, as a
consequence of the local gauge symmetry: the expectation value of any
operator not invariant under local gauge symmetries vanishes, a result
called Elitzur's theorem. \cite{Kogut1979} The magnetization is a
one-body observable, clearly not invariant under the 4-spin flipping
gauge transformation, and it consequently vanishes. The nature of the
low-temperature phase is unveiled instead by the gauge-invariant
correlation function, or the expectation value of bunches of spins
whose positions draw a planar close contour in the lattice. Such a
non-local operator is helpful to interpret the phase transition as a
confinement-deconfinement condensation of kinks, rather than a usual
order-disorder transition found in pairwise models.  In a different context, classical Ising models with four-body
 interactions are also studied as effective models for the interaction of superconducting electrons or grains.\cite{Moore2004,Xu2004}

 On the other hand, different $p=4$ Ising models have been studied
 from a statistical physical point of
 view.\cite{Mouritsen1983,Bouchaud1994,Lipowsky1997,Lipowsky2000,Lipowsky2000bis,Lipowsky2000bisbis,Nishiyama2004}
 They are, in particular, subject of interest as far as their
 plaquette version may exhibit slow dynamics and other dynamical
 features reminiscent to those of glasses, which are self-induced
 (i.e., not induced by quenched disordered
 couplings).\cite{Lipowsky1997,Lipowsky2000,Lipowsky2000bis,Lipowsky2000bisbis}. The
 system with interacting quadruplets defined as the plaquettes of a
 hyper-cubic lattice (the {\it plaquette Ising model}) has been
 particularly studied.  In two dimensions it presents a phase
 transition with dynamical activated behavior.\cite{Jack2005} In three
 dimensions the model is called the Gonihedric
 model,\cite{Ambartzumian1992,Savvidy1994} and is known to exhibit a
 first-order phase transition, and a degenerated ground
 state.\cite{Mouritsen1983,Lipowsky1997,Espriu1997} The slow dynamics,
 metastability and glass-like features of the 3D model have been
 studied in
 Refs. \onlinecite{Lipowsky1997,Lipowsky2000bisbis,Swift2000,Dimopopulos2002}. A
 anisotropic variant of the Gonihedric model has been recently
 studied,\cite{Castelnovo2010} its dynamical properties are shown to
 be signaled by the expectation values of quantum
 information-theoretical estimates in its quantum counterpart.

 The $O(2)$ generalization of the lattice gauge-invariant model,
 called Abelian gauge theory, presents a larger, $O(2)$, local gauge
 invariance. Indeed, its behavior at low temperature is described in
 the continuum limit with the Euclidean action of electrodynamics,
 according to a spin-wave approximation resembling the one allowing to
 describe the undercritical temperature of the $d=2$ $O(2)$ model in
 terms of a Gaussian theory.\cite{Kogut1979} As in the Ising gauge
 theory, the $d=2$ Abelian gauge theory presents no phase transition,
 while the $d=3$ presents a phase transition separating two
 unmagnetized phases and, again, the order parameter being a nonlocal
 contour correlator, an object which is directly related with the
 potential energy of deconfinement, in the field theoretical language.

\section{The Leading models:  $p=4$ XY and Complex Spherical (CSM) models  }
\label{sec:models}

\subsection{Definition of the model}

We are interested in the statistical analysis of the mode wave
interaction Hamiltonian, introduced in section \ref{sec:modelocking}.  
We will restrict our analysis to the four-body interaction term, as it
contains the essential nonlinear phenomenology. The inclusion of the
local interaction due to a non-flat gain, see
Eq. (\ref{eq:master_equation_haus}), does not change the thermodynamic
features of the model, and its inclusion is discussed in Sec. \ref{sec:gain}.

We, then, consider a set of $N$ electromagnetic modes whose amplitudes
are described by the complex numbers $a_m$, $m=1,\ldots,N$, with
phases $\phi_m=\arg a_m$ and moduli $A_m=|a_m|$.  The Hamiltonian, ${\cal H}$, is
completely specified in this case by the the list of {\it quadruplets}, 
or \emph{ordered} sets of four mode indices $(spqr)$, which
correspond to different terms in ${\cal H}$. The list of quadruplets
can be specified by the {\it Adjacency Tensor}, ${\cal A}_{sprq}$,
equal to 1 whenever the quadruplet defined by its indices is a term of
the Hamiltonian, and zero otherwise. Hence, $\cal H$ takes the form\cite{Gordon2002} (see Eq. (\ref{eq:master_equation_haus})):
\barray
{\cal H} = -  \frac{J_0}{8} \sum_{s,p,q,r} {\cal A}_{spqr}  \, A_{s} A_{p} A_{q} A_{r} \nonumber  \\ 
\cos (\phi_{ s}-\phi_{ p}+\phi_{ q}-\phi_{ r} ) 
\label{eq:H4new}
,
\earray
while the mode amplitudes are constraint by Eq. (\ref{sphericalconstraint}). 
This model corresponds to the (ferromagnetic) 4-body Complex Spherical Model (CSM) in
an arbitrary topology of quadruplets. 
In the particular case where the moduli $A_m$ are 
\emph{fixed} and all equal to $1$, the Hamiltonian reduces to the 4-body
XY ($O(2)$) model:
\barray
{\cal H}_{\rm XY} = -  \frac{J_0}{8} \sum_{s,p,q,r} {\cal A}_{spqr}  \cos (\phi_{ s}-\phi_{ p}+\phi_{ q}-\phi_{ r} )
\label{eq:HXY}
.
\earray

\subsection{Symmetry of the list of quadruplets}
\label{sec:symmetry}

The adjacency tensor $\cal A$ is in general not symmetric under
permutations of its indices. 
However 
it exhibits a symmetry which is also in each one
of the terms in the Hamiltonian (\ref{eq:H4new}).  Given an {\it
ordered} set of four indices, its $24$ possible permutations 
(i.e., quadruplets) can be split into $3$ non-equivalent subsets of the $8$
permutations that have the same energy.  Moreover, if a quadruplet
respects the FMC, then all its $8$ equivalent permutations do. The three
non-equivalent permutations can be chosen to be ${\cal
  Q}=\{(1234),(1324),(4231)\}$, in such a way that the Hamiltonian can
be then written as:
\barray
{\cal H} = -  J_0 \sum_{s<p<q<r}\, \sum_{\pi \in \cal Q} {\cal A}_{\pi_s \pi_p \pi_q \pi_r}  \, A_{\pi_s} A_{\pi_p} A_{\pi_q} A_{\pi_r} \nonumber  \\ 
 \cos (\phi_{\pi_s}-\phi_{\pi_p}+\phi_{\pi_q}-\phi_{\pi_r} )
\label{eq:H4alt}
,
\earray
where $\pi_s$ are the members of the permutation,
$\pi=(\pi_1\pi_2\pi_3\pi_4)$. This is the origin of the $1/8$ factor
in Eq. (\ref{eq:H4new}).  The size scaling of $J_0$ is to be fixed in
such a way that the energy $E=\<H\>$ is an extensive quantity in both
low and high-temperature phases. We will treat this point in
Sec. \ref{sec:threshold}.


\subsection{Topology of quadruplets}

In the following analysis we have considered two types of topologies
$\cal A$: 

{\bf 1) Homogeneous Topology (HT).} The quadruplets are selected uniformly at random.
The desired number of quadruplets (or ordered sets of
four indices), $\nq$, are randomly chosen among all the possible quadruplets.
Specifically, 
in order to preserve the permutation symmetry of the Hamiltonian (cf. sec \ref{sec:symmetry}),
this random selection is performed randomly selecting
$\nq/8$ quadruplets among all possible
$N(N-1)(N-2)(N-3)/8$ quadruplets with different energy,
and, for each one, we append all their 8 equivalent permutations to the list. 

We will call {\it fully connected} the particular case of the HT such that all
quadruplets are considered. 

We stress that in the HT case, hence, the list of quadruplets is not conditioned by the set of frequencies. 
In the photonic language, this situation corresponds to the case of the so-called {\it narrow frequency distribution}, 
in which the different frequencies $\omega_n$ are all similar in magnitude, the difference
between them being lower than the linewidth $\gamma$, so that the FMC Eq. (\ref{eq:FMC}) is
trivially satisfied.

{\bf 2) Correlated Topology (CT)}. The quadruplets are
no longer chosen in an uncorrelated way, although still stochastically chosen. 
We randomly select $\nq/8$ quadruplets with different energy, \emph{only} among the possible $\sim N^3$ quadruplets $spqr$ satisfying the relation:
\beq
s-p+q-r=0
\label{eq:myFMC} \, ,
\eeq
and, for each one, we append all its 8 equivalent permutations to the list. 
This prescription is the result of imposing a FMC,
cf. Eq. (\ref{eq:FMC}), if one supposes a set of $N$ frequencies
distributed as a linear comb,
\beq
\omega_m=\omega_0 + m\, \delta\omega
\label{eq:freqs}
, \qquad \delta \omega \gg \gamma, 
\eeq
which is the case of interest describing closed cavity lasers. The FMC
identity has become an integer identity since, in the optical
interpretation, the values $\omega_m$ are to be understood as the
centers of the bins of a discrete frequency distribution whose bin width
is given by the line-width $\gamma$, so that Eq. (\ref{eq:FMC})
becomes equivalent to Eq. (\ref{eq:myFMC}).

Besides having a clear
physical motivation (the equispaced frequency case), the constraint
Eq. (\ref{eq:myFMC}) is also the simpler and most natural way of
introducing correlations in an abstract stochastic set of interacting
quadruplets. Consider the analogy with a random network: a way to
construct random but correlated graphs is introducing some kind of
distance between different nodes (as the absolute value of the
difference between the node indices ${\rm d}_{sp}=|s-p|$), and
choosing bonds with a probability depending of such a distance. 
In the case of the list of quadruplets, one needs a four-index function, 
and a similar role can be played by ${\rm d}_{spqr}=|s-p+q-r|$.
The FMC with the equispaced set of frequencies is equivalent to choosing quadruplets
presenting the minimum value, ${\rm d}_{spqr}=0$. In this way,
the mode frequencies are not a degree of freedom, but a coordinate
driving correlations as distance in a graph.

\vspace{0.5cm}

\begin{figure}[t!]                        
\begin{center} 
 \includegraphics[width=.45\columnwidth]{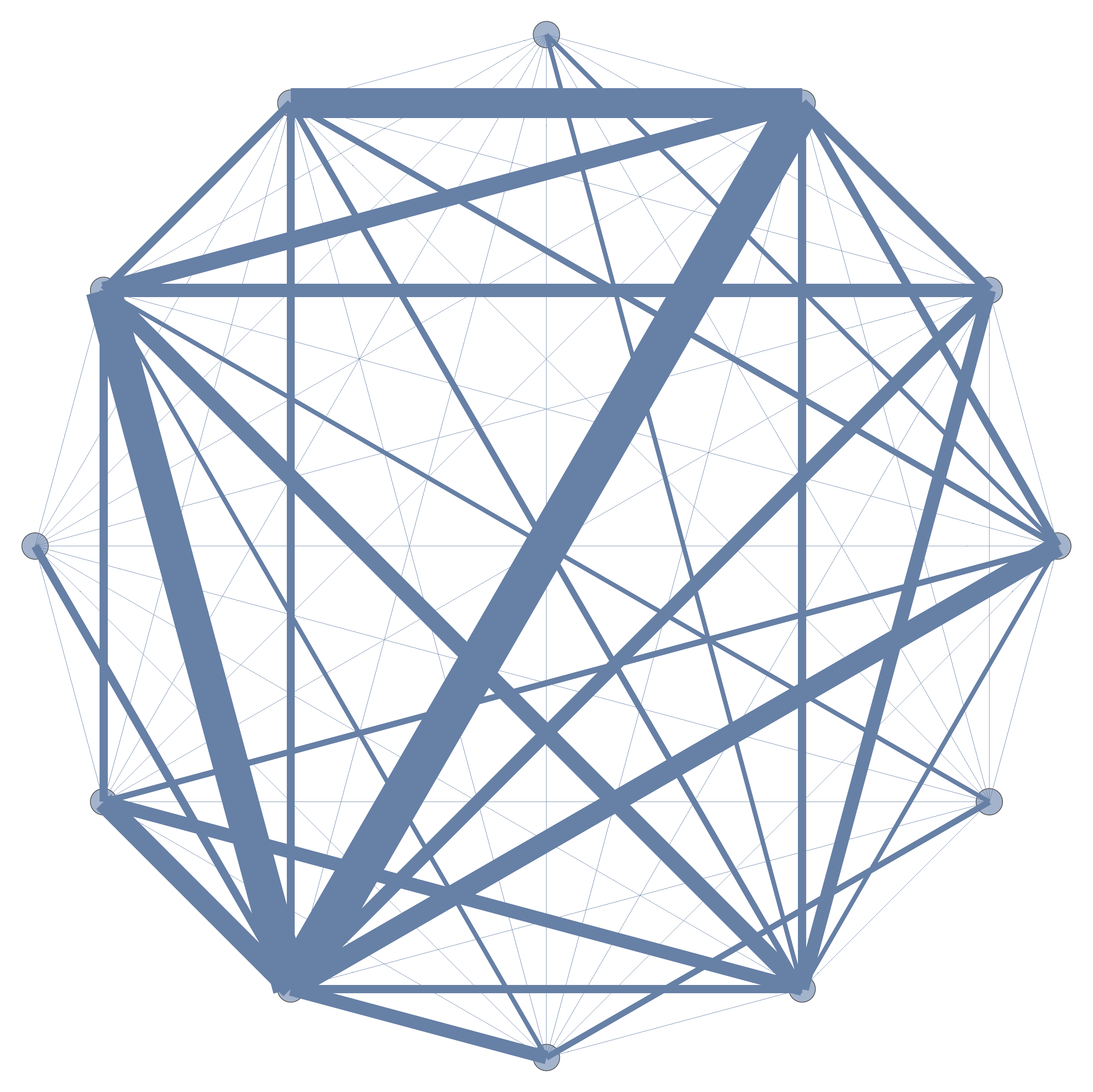}  
 \includegraphics[width=.45\columnwidth]{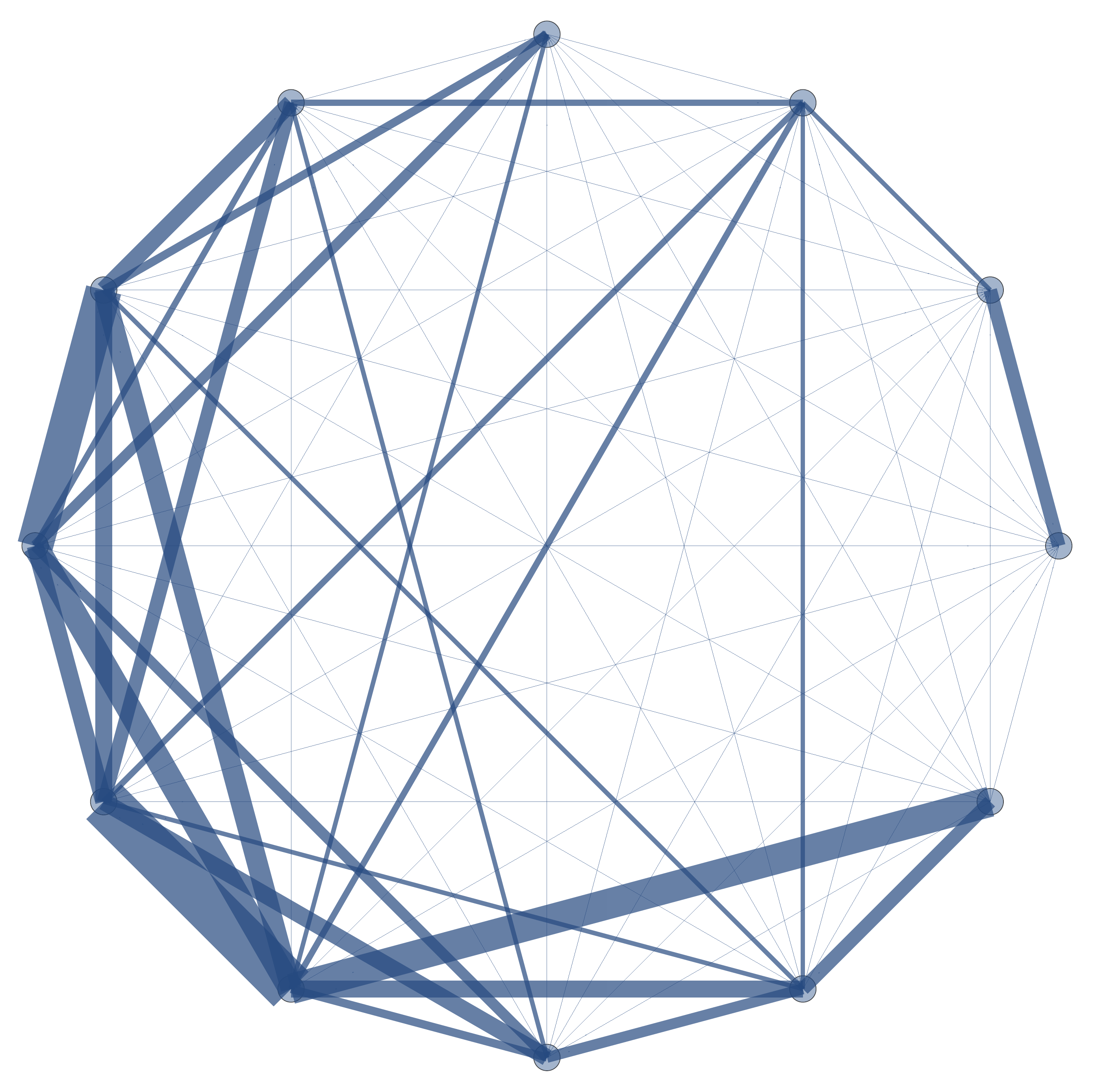}  
\caption{From--quadruplet graphs of two lists of quadruplets with
  $N=12$, $\nq=97$ and Homogeneous (left) and Correlated (right)
  Topology, respectively.  Adjacent nodes represent adjacent mode
  indices, and the thickness of a link is proportional to the number
  of quadruplets that contain the two linked nodes. The angular
  position of nodes corresponds to their index, so that nodes on
  adjacent clock hours have adjacent frequencies. The central
  frequencies are at hour 9 and 10 o'clock. For the HT the thickness is
  uncorrelated to the position. For the CT it is apparent that the
  thickest links are between adjacent modes. Moreover, in the CT case
  the modes at the center of the spectrum (at hour 9 and 10 o'clock)
  share more quadruplets than those at the edge (at hour 3 and 4
  o'clock).  }
\label{fig:quadruplets}
\end{center}   
\end{figure}

While there is a stochasticity in both Homogeneous and CT, due to the
fact that only a random fraction of the possible quadruplets are considered,
there is an important difference: in the HT, the average number of
quadruplets connecting two nodes is independent of the nodes in the
quadruplet, while in the CT one can show that the number of
quadruplets (normalized as a probability distribution) connecting
couples of nodes with frequency difference $\omega_i-\omega_j$, i.e., at a
distance $|i-j|$, is $h(x)=2(-x+1)$ where $x=|i-j|/N$. Modes with
similar frequencies are connected by a higher number of quadruplets
(and, consequently, effectively more coupled) in Eq. 
(\ref{eq:H4new}). This difference is illustrated in
Fig. \ref{fig:quadruplets}: we show the difference through the so
called {\it From-Quadruplet Graph}, or a weighted graph such that
each node represents a mode, and the edge weight (represented by the
line thickness) is proportional to the number of quadruplets
containing their modes.

There are good reasons to classify the interaction topology in the two
types, HT, CT. As we will explain below, if $N_4 \sim {\cal O}(N^{\geq 2})$,
 the thermodynamic behavior of the
system is completely determined by the type of topology of
quadruplets and not by $\nq$, and it is essentially different in the HT
and CT cases. The frequency correlations in the last case induce
correlations between mode amplitudes with different frequencies, $a_j
a_{j'}$, that will drastically modify the thermodynamic phases, as
will be shown in Sec. \ref{sec:resultsSM}. 
%
We stress that there is, then, also a convenience for studying stochastic sets of
quadruplets: as we will discuss, such dilute systems can be
numerically processed more efficiently, hence the usefulness of the
dilute ensemble of quadruplets. This point will be discussed in  Sec. \ref{sec:numerical}.

On the other hand, one may ask why we do study stochastic sets of
quadruplets instead of considering, for example, deterministic sets
given by the four nodes composing a plaquette of a $d$-dimensional
hyper-cubic lattice, as done for the Gonihedric model in the works
already mentioned in the introduction. The answer is given by the fact
that the $p=4$ ferromagnetic Spherical Model, as we explain in the
next section, presents a trivial thermodynamic behavior when the
number of quadruplets is low enough and, in particular, in the
$\nq\sim {\cal O}(N)$ case corresponding to the plaquette-based list
of quadruplets. From the point of view of optics, on the other hand, the
present system is relevant for the description of a closed cavity
laser, such that, in principle, each mode interacts with the rest of
the modes (the fully connected case). This would lead to $\nq \sim
\order(N^4)$ (or to $\nq \sim {\cal O}(N^3)$ with the constraint
Eq. (\ref{eq:myFMC})), a situation which is incompatible with the
plaquette-based topology.

\section{Role of the quadruplet dilution threshold}
\label{sec:threshold}

\subsection{ Non-equipartite condensation in the Spherical Model.}

As will see, the Complex Spherical model presents a trivial
low-temperature behavior, that will be called {\it Non-equipartite
  Condensation}, whenever the number of quadruplets is low enough,
$\nq \sim \mathcal{O}(N^{<2})$ for the ferromagnetic case.  The non-equipartite
condensation is such that all the spherical constraint
Eq. (\ref{sphericalconstraint}) is concentrated in a low, $\order(1)$
number of sites, whose amplitudes are $A\sim \order(\sqrt{N})$. In this case,
the energy in the low-temperature phase is of order $E\sim -J_0 N^2$. 
The {\it Equipartition}, alternative to the non-equipartite condensation, 
is characterized by a $A\sim\order(1)$ in both phases, hence $E\sim -J_0 N_4$. 
In the latter case, the
low-$T$ phase is characterized by the homogeneity of spin moduli, which
tend to {\it lock}, i.e., to become equal throughout the system,
contrarily to the former case.  One observes that the energy is lower
(of a larger order with $N$) in the non-equipartite condensation
whenever $\nq \sim \mathcal{O}(N^{<2})$. Requiring the extensivity of the energy one obtains that, according to the type of
condensation, $J_0$ is subject to satisfy the following scaling
\beq J_0 \sim \left\{
\begin{array}{lll}
1/N & \mathrm{non-equipartition} & (N_4 \sim \mathcal{O}(N^{<2})) \\
N/\nq & {\rm equipartition} & (N_4 \sim \mathcal{O}(N^{>2}))
\end{array}
\label{eq:J0scaling}
\right.
.
\eeq

For high enough temperature, one expects a disordered phase with
uncorrelated and equipartited spins. The extensivity 
of the energy requires $J_0 \nq \sim \order(N)$,
implying in its turn that the non-equipartite condensation do not occur for
$\nq \sim \order(N^{ > 2})$, confirming Eq. (\ref{eq:J0scaling}). 

This argument does not apply to the marginal situation $\nq \sim \order(N^2)$. We expect, however, equipartition, since in this circumstance there is an extensive entropic contribution to the  free energy. 
This is in agreement with our numerical results for all the considered systems satisfying $\nq \sim \order(N^2)$, which turn to be equipartite.


In the following we are interested in the equipartite case.
We, hence, consider from now on systems with $\nq \sim \mathcal{O}(N^{\geq 2})$. Our
Hamiltonian, in its final form, will be taken as (see
Eq. (\ref{eq:J0scaling})):
\barray
{\cal H} = -  \frac{N}{8\nq} \sum_{spqr}\ {\cal A}_{s p q r}  \, A_{s} A_{p} A_{q} A_{r}  \nonumber  \\ 
 \cos (\phi_{s}-\phi_{p}+\phi_{q}-\phi_{r} )
\label{eq:H4def}
.
\earray

\subsection{Non-equipartition in the Disordered Spherical Model.} 
Although in the next chapter our numerical analysis focuses on the
ferromagnetic case, for completeness we also discuss how the
non-equipartite condensation occurs in the
quenched disordered case below the higher threshold $\nq \sim \mathcal{O} (N^{ 3})$. 
The argument is based on a mean field approximation
allowing to compute the scaling of the average energy with $N$, $\nq$
within the replica formalism.  The details can be found in the
Appendix \ref{app:energy}. Supposing that the coupling $J$ in
Eq. (\ref{eq:H4new}) is no longer ferromagnetic but Gaussian
distributed with average $J_0$ and variance $\sigma$, one has that the
energy scaling in both non- and equipartite types of condensation goes
as:
\beq E \sim \left\{
\begin{array}{ll}
\mathrm{non-equipartite} & -(J_0+\sigma)N^2 \\
{\rm equipartite} & -(J_0+\sigma^2)N^2
\end{array}
\label{eq:Escalingdisorder}
\right.  , \eeq 
so that for the extensivity of the energy, one is
forced to take for $\sigma$ the minimum between $N/\nq$ and
$1/N^2$. Hence, the threshold between non- and equipartition
becomes in this case $\nq = \order(N^3)$. 
This threshold is compatible with the provisional results of our
simulations in the presence of disorder (that will be reported in a
future communication).

\subsection{Magnetized-to-unmagnetized threshold of the XY model for low number of quadruplets.} 

An equivalent threshold effect is observed for the $p=4$ XY ferromagnet, Eq. (\ref{eq:HXY}), with HT.
In this case the threshold is, instead, the extensive situation $\nq \sim \mathcal{O}(N)$,
above which the system presents a low temperature phase with spontaneous breaking of
the $O(2)$ symmetry. Below and at the threshold, i.e., for $\nq \sim \mathcal{O}(N^{\leq 1})$, the model
remains unmagnetized. This fact will be discussed in more detail in
Sec. \ref{sec:resultsXY}.


\section{Numerical analysis}
\label{sec:numerical}

\subsection{Efficient Monte Carlo simulation: 
the synchronous Monte Carlo algorithm}

We have performed a Monte Carlo (MC) integration using a home-made
algorithm dealing with vector $p=4$ interaction models in arbitrary
topologies. The algorithm uses local updates (in the case of the
Spherical Model it is not possible to use cluster updating, due to the
non-locality induced by the spherical constraint).  The Parallel
Tempering algorithm has been used to enhance equilibration in large
systems.

Moreover, for most of the results presented in this article we have
used a parallel, high-performing version of the algorithm, running on
Graphics Processing Units. The parallel Monte Carlo integration of a
system of interacting spins requires the division of the set of spins
in non-interacting subsets, such that the members of each one can be
processed in parallel.  In bipartite lattices, such a division is
called the checkerboard decomposition, while in general graphs
defining the pairwise interaction it is necessary to perform the {\it
  coloring} of the graph, in such a way that all spins with equal
color are processed in parallel, and different colors are processed
sequentially.\cite{Berganza2013} As explained before, the case of
interest is a system in which the topology of the interaction is given
by a set of at least $\order (N^2)$ quadruplets between $N$ modes, so that each
mode possesses an extensive number of {\it quadruplet neighbors},
i.e., of modes such that there is at least a quadruplet connecting
both. As a consequence, the MC parallelization of such a kind of
highly connected system is, in principle, unfeasible.

However, parallel Monte Carlo techniques can still be used
in this case. We have observed that, quite remarkably, there are
circumstances (that will be specified elsewhere) in which applying the
so-called {\it Synchronous Monte Carlo} rule (i.e., to all spins in
parallel, regardless of their connectivity), one recovers the correct
results. Although one is making an error in each update (since one
updates interacting spins simultaneously), the overall error averages
down to zero.  In a fully connected $p=2$ spin model the
fully-parallel MC update does not differ with respect to a sequential
MC scheme, as it has been already observed.\cite{Nobre2001} In the
present case with $p=4$ body interactions and $\mathcal{O} (N^{\geq 2})$ quadruplets,
the results of the Synchronous Monte Carlo are, again,
consistent from the serial algorithm.

It is particularly remarkable that for the present model it is not necessary to
have a fully connected system for the Synchronous Monte Carlo
algorithm to work: a dilute, but connected enough system is sufficient
to obtain results which are indistinguishable from that of the serial
MC algorithm. Interestingly, this holds true even if the transition is
no longer describable in mean field approximation: we will show that
in the CT case fluctuations arise and change the nature of the
transition and, even in this case, the Synchronous Monte Carlo leads to
correct results.

\begin{figure}[t!] 
\begin{center} 
\includegraphics[width=1.\columnwidth]{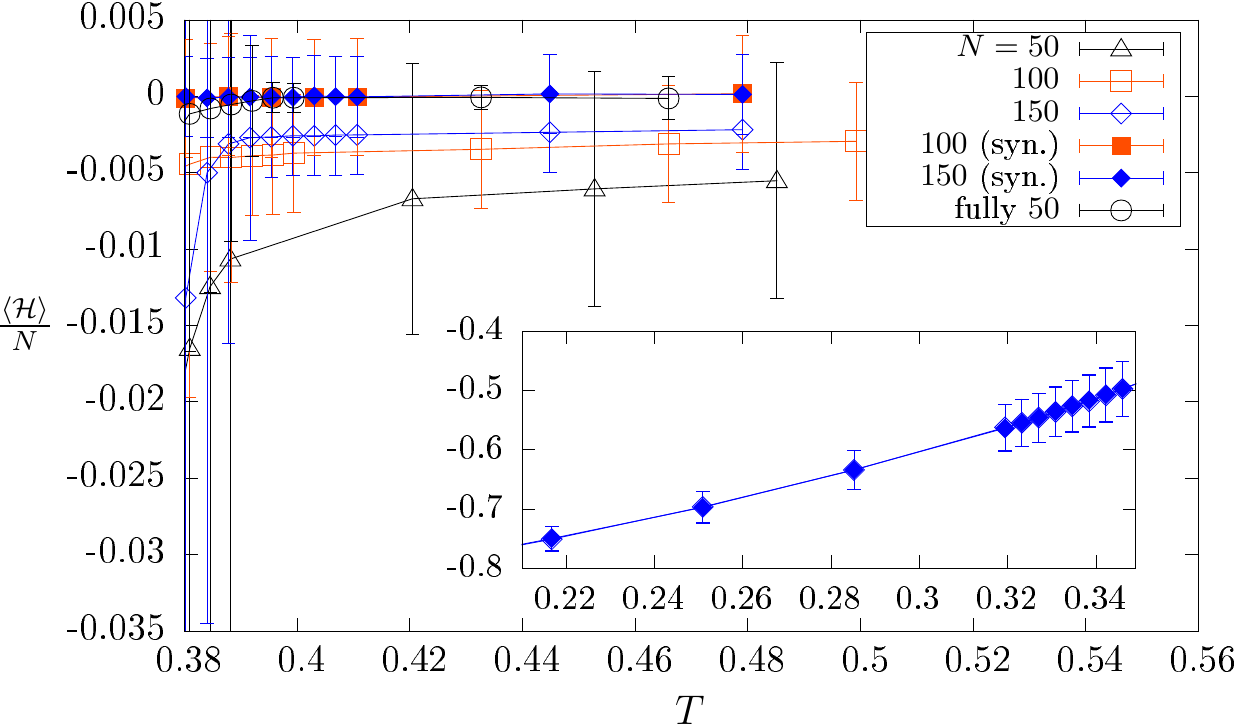}
\caption{High temperature intensive energy versus temperature for systems with three sizes in a
  diluted HT with $\nq = N^2$ quadruplets, computed with the serial MC
  algorithm (open symbols for $N=50,100,150$). The results obtained with
  the synchronous update for these systems (corresponding full symbols
  for $N=100,150$) yield an average value closer to zero, though compatible in the statistical uncertainty.
  At low temperature both algorithms accurately coincide and the results are indistinguishable, as displayed in the inset for $N=150$. 
  Note as the fully connected system (open circles), simulated by means of a
  serial MC exhibits a zero energy at high $T$ yet for $N=50$. 
}
\label{fig:comparisonNrg}
\end{center}
\end{figure}

An example of the reliability of the Synchronous Monte Carlo is shown in Fig. \ref{fig:comparisonNrg}
for the average energy in the HT, but the picture is valid also for the CT case.
The measures are always compatible for a serial MC and a parallel MC.
In the low temperature phase, in particular, the values are numerically indistinguishable.
Some appreciable deviations are observed only in the high temperature phase in the case of diluted systems ($\nq = N^2$ in Fig. \ref{fig:comparisonNrg}).
In this case the synchronous algorithm predicts an average energy which is closer to zero,
although compatible with the serial algorithm within thermal fluctuations.
For $T>T_c$, the synchronous algorithm has hence the effect of masking finite size effects, since the energy 
per site $E/N$ vanishes at large $N$ for $T>T_c$.

\subsection{Observables of interest}

Besides the energy $E=\<{\cal H}\>$, we will consider the following
observables. Firstly, the specific heat:
\barray
c = \frac{1}{N} \frac{\partial \<{\cal H}\>}{\partial T}  = 
\frac{\<{\cal H}^2\>-\<{\cal H}\>^2}{N\,T^2} .
\earray
Also, the {\it average modulus}, $\<r\>$ with $r$ being
%
\beq
r = \frac{1}{N}   \sum_j A_j  
\eeq
a quantity which is related with the site-fluctuations of the modulus:
$\frac{1}{N}\sum_{j}\, (|a_j|-r)^2 = 1-r^2$: the larger $r$, the more
{\it locked} are the moduli of the spins in a given configuration, and
$r=1$ corresponds to a configuration with all the mode amplitudes that
have modulus equals to one. 
Another interesting observable is the {\it magnetization}, $\<m\>$, where $m$ is the complex number:
\beq
m = \frac{1}{N} \sum_j a_j
,
\eeq
along with its Cartesian components, $m_x={\rm Re}[m]$, $m_y={\rm Im}[m]$. 
%

Finally, we also measure {\it frequency correlation functions}, which
are ensemble averaged correlations between modes whose frequencies
differ by a given quantity $\omega$. These observables
acquire full sense in the CT case, when the mode frequencies play a
role in the topology and, hence, in the thermodynamics. We define, in
particular, the {\it intensity correlation function} $C_{\rm i}$:
\beq
C_{\rm i}(\omega) = \frac{1}{K(\omega)}\, \< \sum_{i,j=1}^N  A_i^2\,A_j^2\, \delta(\omega_i
-\omega_j+\omega)  \>
\label{eq:Cs}
\eeq
$K(\omega)=\sum_i \sum_j \delta(\omega_i-\omega_j+\omega)$ being the normalization, along with the {\it phase correlation function},
$C_{\rm p}$:
\beq C_{\rm p}(\omega) = \frac{1}{K(\omega)}\, \< \sum_{i,j=1}^N
\cos(\phi_i-\phi_j)\, \delta(\omega_i-\omega_j+\omega) \>
\label{eq:Cp}
.
\eeq
We also define their respective {\it connected} functions: \beq \bar
C_{\rm i}(\omega) = C_{\rm i}(\omega) - \frac{1}{K(\omega)}\, \sum_{i,j=1}^N \< A_i^2\>\,
\<A_j^2\>\, \delta(\omega_i-\omega_j+\omega)
\label{eq:cCs}
\eeq
and idem for $\bar C_{\rm p}(\omega)$.

\subsection{Details of the simulations}

We have considered finite-size realizations for several values of $N$, ranging from  $N=50$ to $N=10^3$, 
depending on the topology and on $\nq$. As an equilibration test we have verified the stationarity of the
distributions of observables in different Monte Carlo time windows of
exponentially increasing length, and the symmetry of the histograms of
the single components of the magnetization, $h(m_{x,y})=h(-m_{x,y})$
(cf. Fig. \ref{fig:maghis}).

\begin{figure}[]                        
\begin{center} 
 \includegraphics[width=1.\columnwidth]{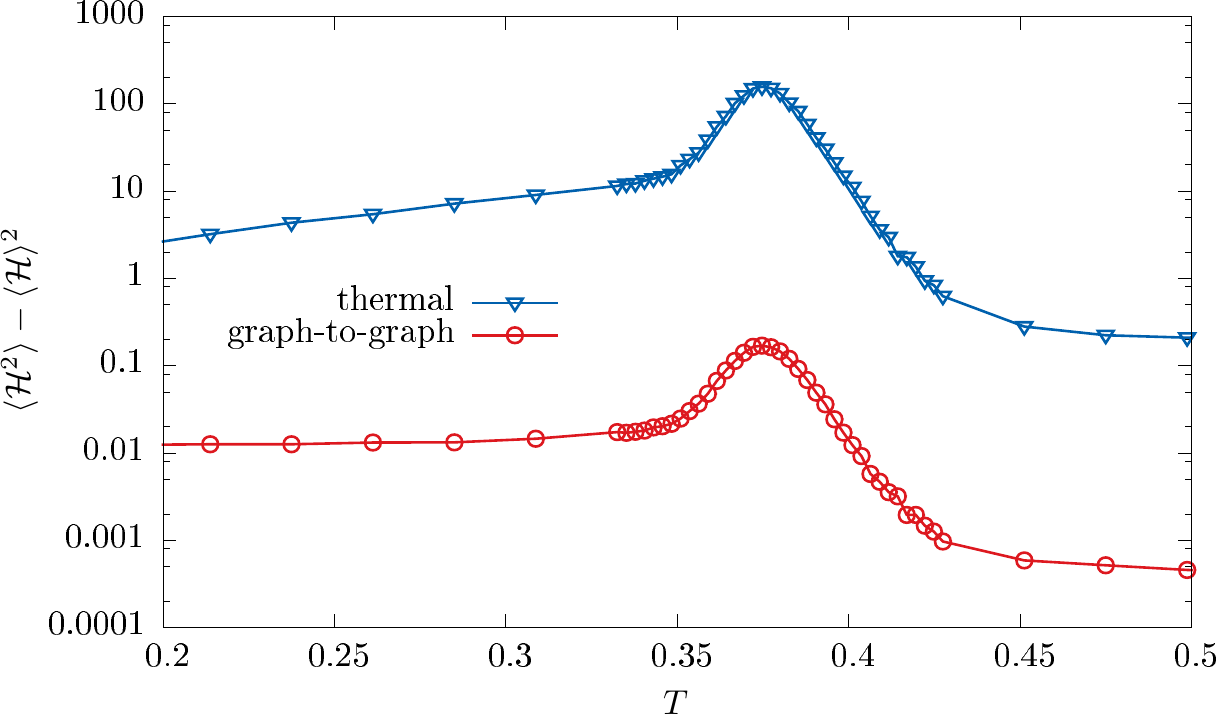}  
\caption{Thermal energy fluctuations versus topological
  (graph-to-graph) fluctuations of a CT system with $N=50$ and with $\nq = N^{2}$ quadruplets.
  The topological fluctuations have been computed over a set of
  100 realizations of the list of quadruplets.}
\label{fig:graphReals}
\end{center}   
\end{figure}

Throughout our analysis, we
have not performed systematic averages over realizations of the
list of quadruplets, in none of the topology types (HT, CT). This is
justified since the fluctuations of thermodynamic quantities among
different realizations of the interaction network are at least one
order of magnitude less than thermal fluctuations. In
Fig. \ref{fig:graphReals} we show how thermal fluctuations are larger than
topological fluctuations of the energy in the worst case analyzed: the
$N=50$ with $\nq\sim \mathcal{O} (N^{2})$ quadruplets distributed with the CT (i.e.,
the smallest, most inhomogeneous system).

\section{Numerical results for the Complex Spherical Model}
\label{sec:resultsSM}

We now present the results of our Monte Carlo analysis for the Complex
Spherical Model. The most salient feature of our simulations is the
presence of a phase transition of first-order nature. The phase
transition separates a high-$T$ phase with randomly distributed
degrees of freedom, zero magnetization and zero energy per mode for
large $N$ 
from a low-temperature phase with
\numero{1} {\it locked moduli and phases} 
\numero{2} non-zero spin-spin correlations, at least in single
configurations and for moderate time scales. 

A remarkable observation is the irrelevance of random dilution: for
both kinds of topologies the results of our simulations are
independent on the number of quadruplets, as far as this quantity is
above the threshold $\nq \sim \order (N^{\geq 2})$ corresponding to equipartite systems.
This means that the results (with the only exception of
the finite-size scaling of the critical temperature) remain unchanged
in the broad range of $\nq$ scaling from $\sim N^4$ down to $\sim
N^2$.

There is an essential difference between the thermodynamic behavior of
the system in the presence of Homogeneous and Correlated Topologies:
in the first case the behavior is compatible with the mean-field
solution, the low-temperature phase is spontaneously magnetized, and
spin-spin correlators are nonzero. On the other hand, for CT's the
results significantly differ from the mean field solution; there is
lack of spontaneous magnetization at low temperatures, and two-point
correlators turn out to vanish.

\begin{figure}[t!]                        
\begin{center} 
 \includegraphics[width=1.\columnwidth]{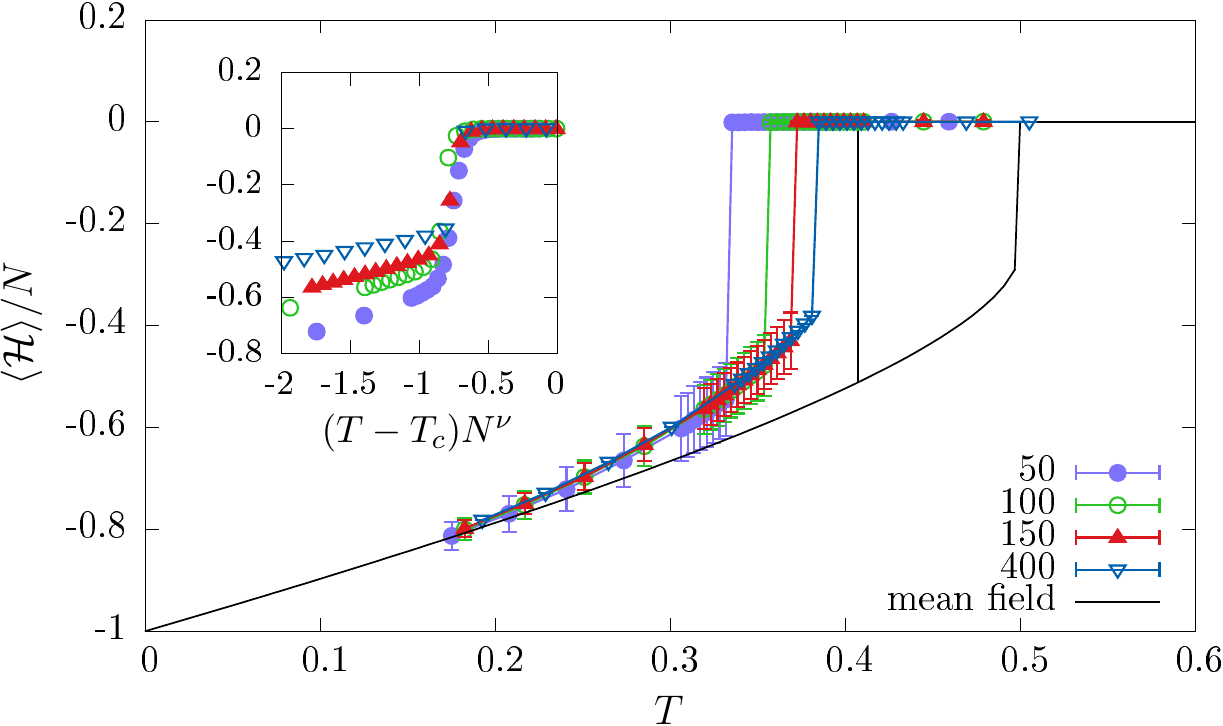}  
\caption{Energy per spin versus temperature for several sizes in the
  HT with $\nq = N^2$. The finite-size critical temperature $T_c(N)$ increases with
  size. The inset shows that the data satisfies a scaling of
  the form $T_c(N)-T_c^{\rm mf}\sim N^{-b}$, with $b=0.6$,
  indicating that they are compatible with the mean field critical
  temperature for large $N$.
  }
\label{fig:nrgHT}
\end{center}   
\end{figure}              

\subsection{Homogeneous topology}

\subsubsection{General features and comparison with mean field theory}

We will first consider the HT system and the fully connected case
(i.e., with all the possible quadruplets active),
 as a particular case of it. Our first, already mentioned result is that,
given the topology type, the dilution turns out to be irrelevant: the
intensive quantities for values of $\nq$ lower than its
maximum value are indistinguishable, within statistical errors, from
that of the fully connected case.
We thus expect the
  behavior in the HT to coincide in the large-$N$ limit with the mean
  field solution of the model,\cite{Gordon2002,Antenucci2014} cf. Appendix \ref{app:meanfield}. 
  This predicts for the
transition temperature $T_c^{\rm mf} = 0.40726$. In
Fig. \ref{fig:nrgHT} we present a finite size analysis of the energy
$E$ in the case of a homogeneous set of $\nq \sim \mathcal{O} (N^2)$ quadruplets. The
high-temperature phase has zero energy, and decreases discontinuously
at a size-dependent value $T_c(N)$. 
As shown in the figure inset for the $\nq \sim \mathcal{O} (N^2)$ case, the
finite size scaling of the transition temperature,
$T_c(N) = k N^{-b}+T_c(\infty)$, leads to an infinite-volume
$T_c(\infty)$ which is compatible with $T_c^{\rm mf}$, for all the
studied sets of quadruplets.\cite{Antenucci2014statistical}

In figure \ref{fig:nrgHT} we also present the energy as a
  function of temperature for the marginal mean field solution shown
  in Appendix \ref{app:meanfield}
\barray
E_{\rm mf}(T)  &=&-\frac{1}{4}[1+(1-2T)^{1/2}]^2 \, , \qquad T<T_c \, .
\label{eq:mfnrg}
\earray

It is apparent how, while the value of the critical temperature is
compatible with the mean field analytical solution, the behavior of
$E_{\rm mf}(T)$ below the transition does not coincide with the numerics.
Indeed, the solution Eq. (\ref{eq:mfnrg}) corresponds to the solution
of two {\em uncoupled} real spherical models with apart spherical
constraints, as it can be seen analytically and verified numerically.  We consider
this observation as the evidence of the inaccuracy of this solution in the generic case
of coupled real and imaginary parts of the complex amplitudes.
To restore the entropy corresponding to the angular degree of freedom
(the extra freedom coming from the global constraint, which is less
restrictive than two independent constraints), it may be necessary to
consider corrections with contributions of $\order(N)$ to the
saddle-point equations, changing the solution (and its
stability). According to this argument, the marginally stable mean
field energy (\ref{eq:mfnrg}) must be lower than the corrected mean
field solution as, indeed, observed in Fig. \ref{fig:nrgHT}.

This problem is absent 
in both the XY and the real spherical model,
for which the respective mean field solutions exactly describe the behavior
of finite size systems already at quite small sizes at all temperatures,
as we have verified numerically.

\begin{figure}[t!]                        
\begin{center} 
 \includegraphics[width=1.\columnwidth]{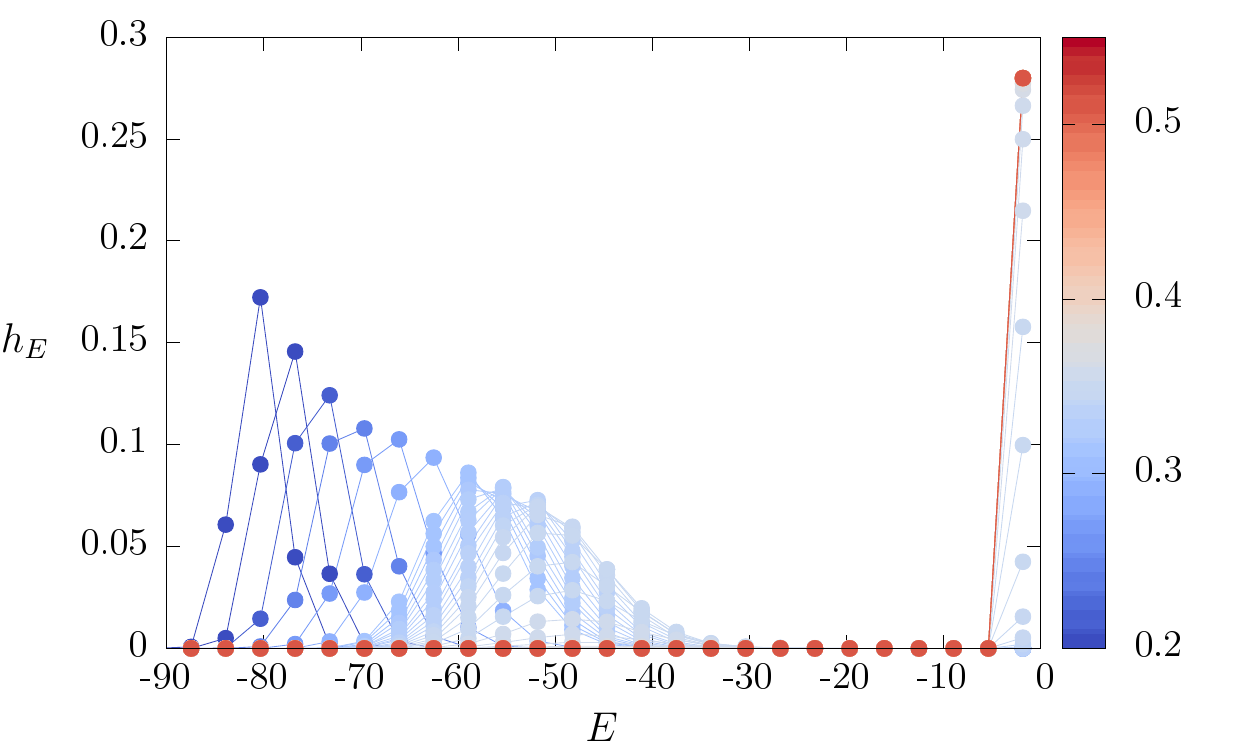}  
\caption{Histogram of the energy at several temperatures of a $N=100$
  system in the HT with $\nq = N^2$. There is a temperature range in which the high-$T$
  and the low-$T$ peaks coexist.
  }
\label{fig:nrghistogram}
\end{center}   
\end{figure}              

\begin{figure}[h!]                        
\begin{center} 
 \includegraphics[width=1.\columnwidth]{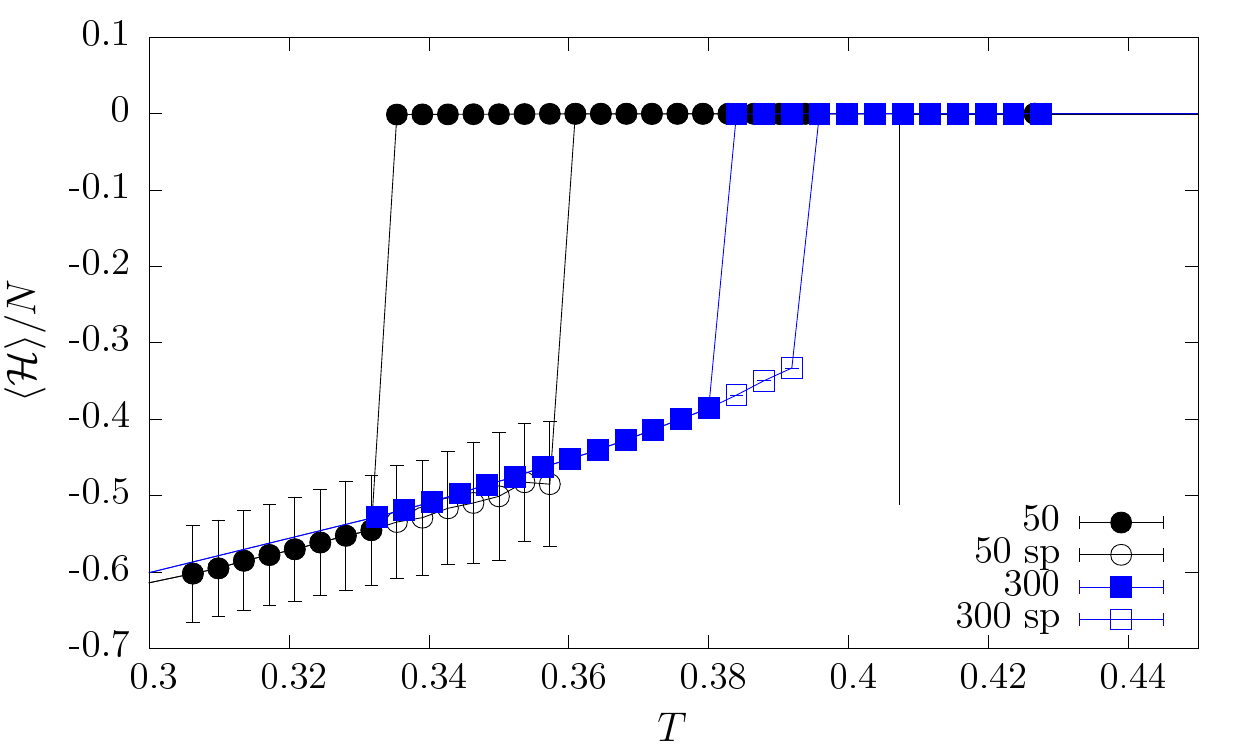}  
\caption{Stable and metastable energy per spin in systems with
  $N=50$, $300$, HTs and  with $\nq = N^2$. The temperature endpoint of metastability
  approaches the finite size critical temperature for large and larger
  sizes.  The vertical line signals the mean field critical point.
  }
\label{fig:nrgspinodal}
\end{center}   
\end{figure}

The finite size transition temperatures $T_c(N)$ reported as vertical
lines in Fig. \ref{fig:nrgHT} have been calculated from the bimodal
energy probability distribution (cf., Fig. \ref{fig:nrghistogram}), as
the temperatures at which the high- and low-energy peaks enclose equal
areas. We have also considered the metastable continuation of the
disordered phase energy, averaging over the disordered peak only, and
a temperature limit of the metastable regime (a {\it Spinodal Temperature} $T_s$) 
as the one at which the low-energy peak vanishes. 
As shown in Fig. \ref{fig:nrgspinodal}, the quantity
$T_s(N)-T_c(N)$ decreases with increasing size, thus indicating that
again the metastability behavior of the model is different from that
predicted by the marginally stable mean field solution, which predicts
$T_s=1/2$ in the thermodynamic limit (the metastable energy being that
of Eq. (\ref{eq:mfnrg}), continued to $T_s=1/2$). The shrinking of the
metastable interval persists even using a Monte Carlo protocol which
favors the relaxation towards the (low-$T$) metastable phase, i.e.,
starting from an ordered configuration and switching off the Parallel
Tempering algorithm.  
The observed decreasing of the metastable interval with system size is so
strong 
that, with the actual statistics and temperature grid, for the largest simulated sizes we are not able 
to observe any spinodal point distinct from the critical point in the statistical error.
 For the largest sizes the low energy peak in the energy distribution associated to the ordered phase, 
 see Fig. \ref{fig:nrghistogram}, disappears  as the zero energy peak of the paramagnetic phase appears.
This situation is different from what
happens in the marginal mean field solution described in Appendix \ref{app:meanfield}. 
We note also that the finite-size nature of metastability in
temperature-driven transitions has already been observed in a
ferromagnetic model with pairwise
interactions.\cite{Berganza2014}

\begin{figure}[t!] 
\begin{center} 
 \includegraphics[width=1.\columnwidth]{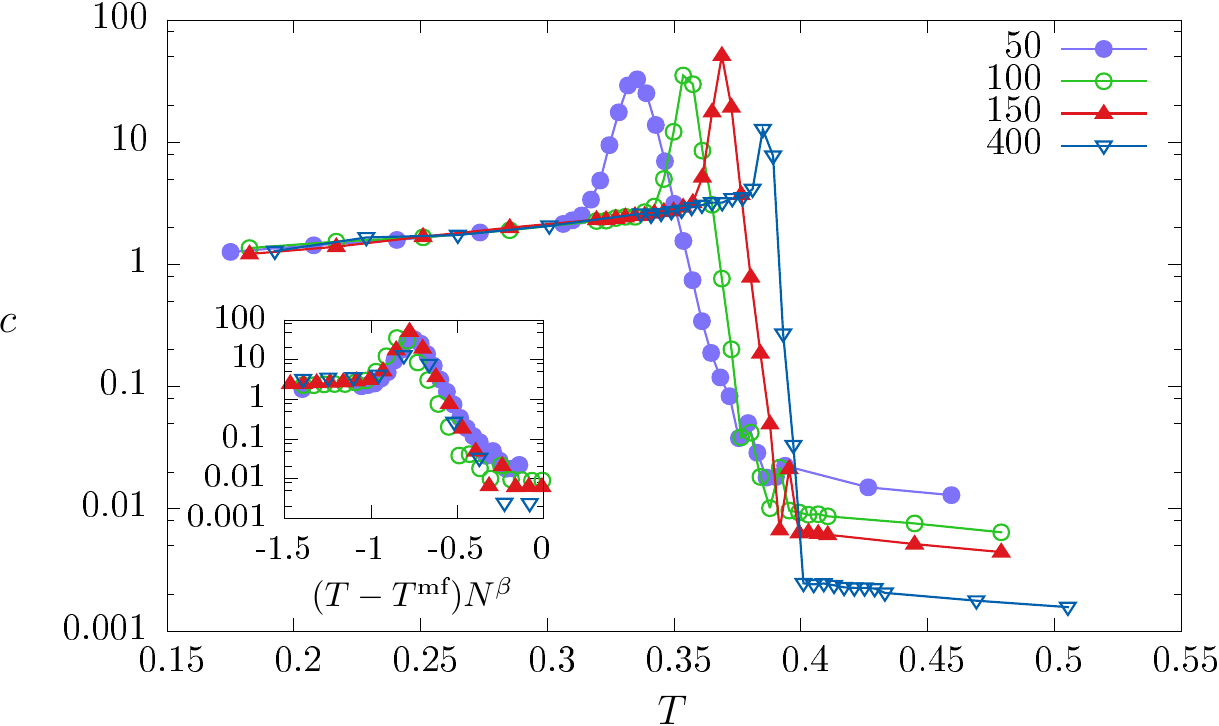}  
\caption{Specific heat versus temperature for different sizes of HT. 
The dataset is the same of figure \ref{fig:nrgHT}. 
The inset shows the data with the scaling relation $T_c(N)-T_c^{\rm mf}\sim N^{-b}$, with $b=0.6$.  
}
\label{fig:cVnoFMC}
\end{center}   
\end{figure}

\begin{figure}[b!]
\begin{center} 
 \includegraphics[width=1.\columnwidth]{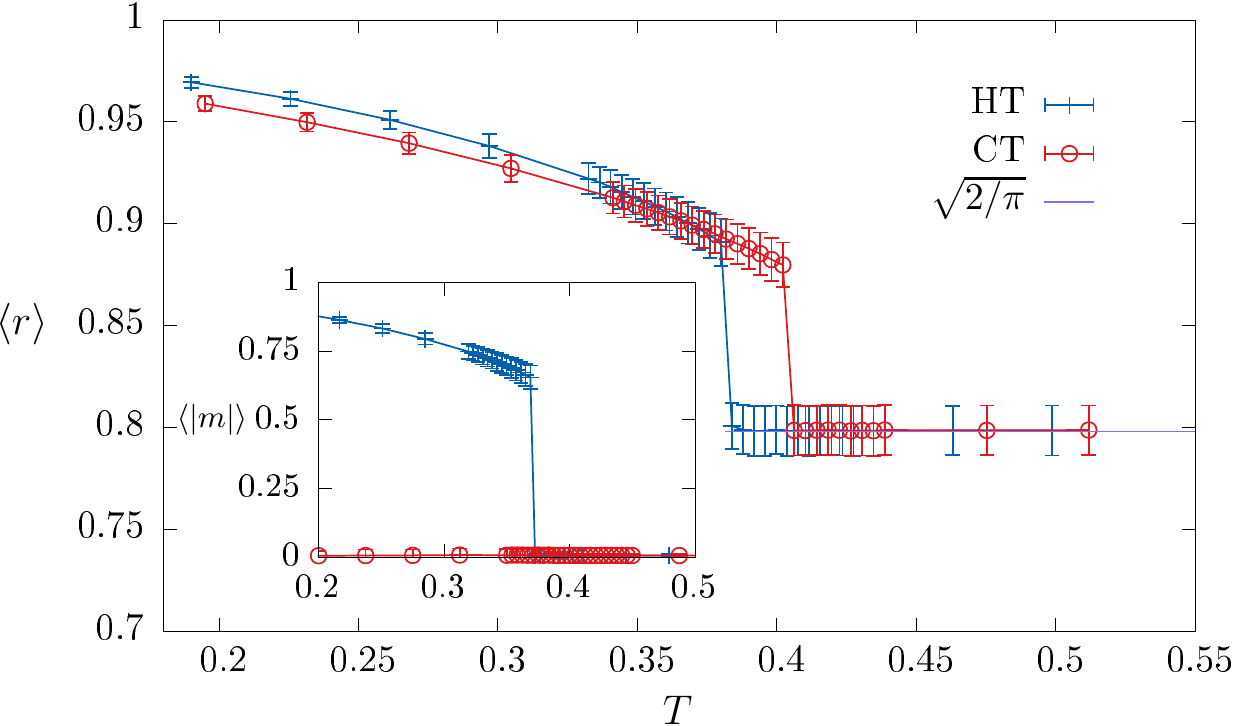}  
 \caption{Average moduli versus temperature for systems with $N=300$,
   HT and CT. Inset: average modulus of the magnetization versus
   temperature for the two systems. 
   The CT system is unmagnetized also at low temperatures. }
\label{fig:rm}
\end{center}   
\end{figure}

The specific heat is presented in Fig. \ref{fig:cVnoFMC}. One observes
no divergence with increasing system size, and a finite size scaling
confirming the one of the energy reported in Fig. \ref{fig:nrgHT}. The
average modulus and the magnetization are presented in
Fig. \ref{fig:rm}. For high temperature, the average modulus achieve
(up to $\order(N^{-1/2})$ fluctuations) the value $(2/\pi)^{1/2}$,
which is the average modulus of uncorrelated complex random variables
satisfying the spherical constraint, as can be exactly proven for large $N$. 
The value of $\<r\>$ is discontinuous at the transition and
converges to $1$ for $T\to 0$, which means that all the spins exhibit
equal modulus, $|a_j|=1$. The magnetization vanishes in the high-$T$
phase and it is $1$ for zero temperature, indicating that it is of
ferromagnetic nature: not only the moduli are locked but also all the
phases coincide  and both phases and moduli lock
at the same temperature, as predicted by mean-field theory. \cite{Antenucci2014}

\subsection{Correlated Topology of quadruplets}

We will now describe the differences induced by the presence
of inhomogeneity of quadruplet topology, due to the FMC.  The
inhomogeneities promote fluctuations on the radial and angular degrees
of freedom, not describable in mean field approximation. As a
consequence, the behavior substantially differs from the HT case.

\subsubsection{Absence of spontaneous $O(2)$ symmetry breaking}

\begin{figure}[t!]                        
\begin{center} 
 \includegraphics[width=1.\columnwidth]{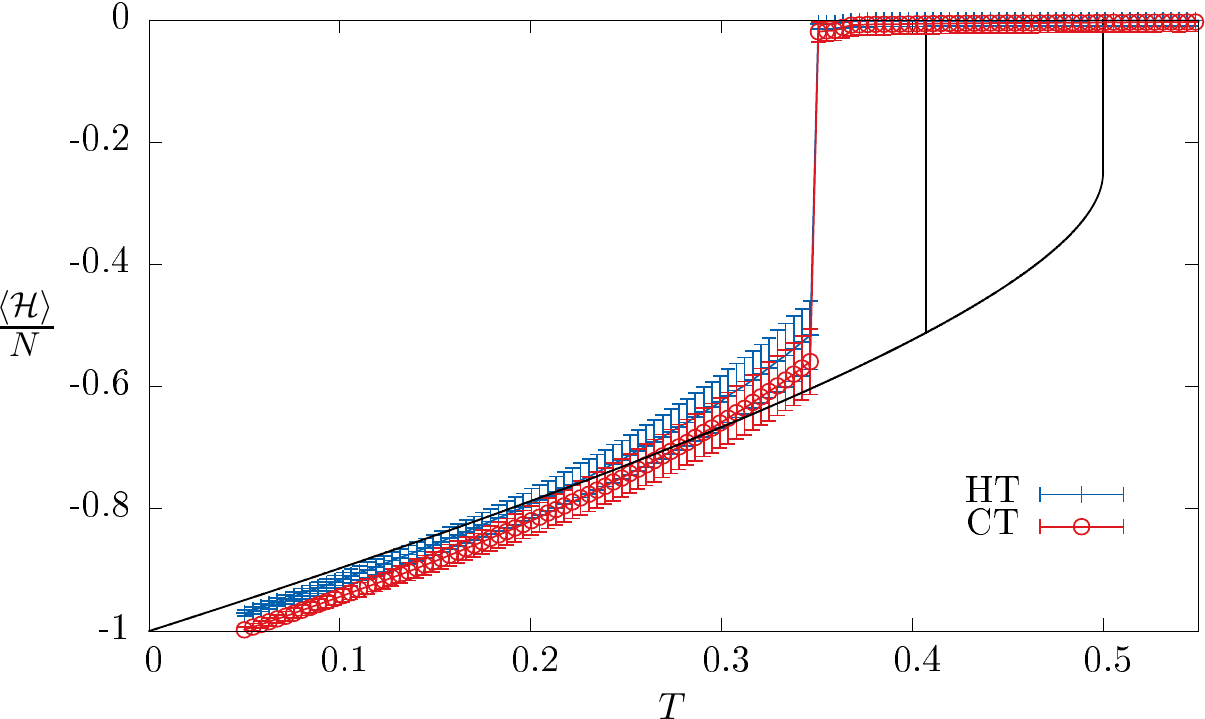}  
\caption{Comparison between energy of systems with Homogeneous and CT with $N=100$ modes and $\nq = N^2$. The black line is the marginal stable mean field solution.
The HT data is observed to converge at $-1$ for $T \to 0$ increasing the size, while the CT remains lower than the HT curve at low temperatures, 
and converges to a value lower than one for $T\to 0$. 
  }
\label{fig:nrgComparison}
\end{center}   
\end{figure}

\begin{figure}[t!]                        
\begin{center} 
 \includegraphics[width=1.\columnwidth]{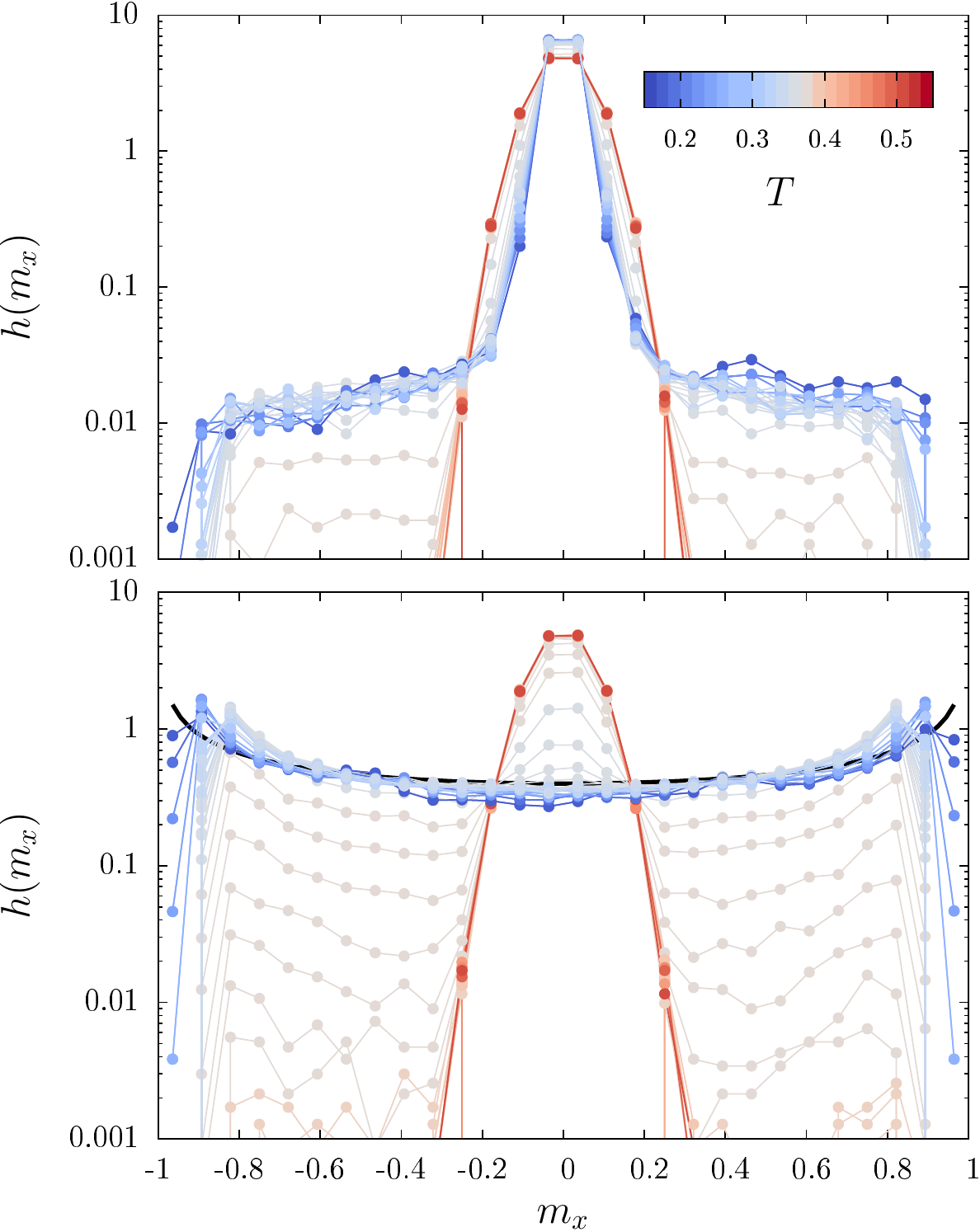}  
\caption{Histograms of a Cartesian component of the magnetization, for
  two systems with $N=300$ and $\nq = N^2$, HT and CT (lower panel and upper panel
  respectively).  The HT case is fully magnetized and for zero
  temperature it converges to the function $(2\pi(1-m^2))^{-1/2}$
  (indicated as a black curve in the lower panel).  The CT system is
  unmagnetized at all temperatures. 
  }
\label{fig:maghis}
\end{center}   
\end{figure}

Although the transition remains first order, and qualitatively equal
to that of the HT case, the energy density stays below the HT case (see
Fig. \ref{fig:nrgComparison}). 
Again, the energy density is independent from $\nq$ up to fluctuations.
 The most dramatic difference induced by the
CT is, however, seen in the average magnetization, which vanishes {\it for all
  temperatures}. We show the change in the magnetization behavior
comparing the histogram of the magnetization components $m_x={\rm Re
}[m]$, Fig. \ref{fig:maghis}. In both HT and CT cases, the
high-temperature phase is unmagnetized with a Gaussian distribution of
$m_x$ centered in zero. In the Homogeneous case, the low temperature
phase is magnetized ($|m|^2\to 1$ for $T\to 0$) following a phase
direction ($\phi={\rm arg}\, m$) which is degenerated, and whose
average projection in the $x$ axis results in the peaks of
$h(m_x)$. For zero temperature, the distribution coincides, indeed,
with $h(m)=(2\pi)^{-1} (1-m^2)^{-1/2}$, or the $m$ distribution
corresponding to a homogeneously distributed $\phi$,
cf. Fig. \ref{fig:maghis}; in other words, the average magnetization
is zero for HT, but this happens since the single configurations are fully
magnetized over an angle which is degenerated. {\it In the CT case, on
  the other hand, the low temperature phase is unmagnetized: the
  average and the most probable magnetization remains zero for
  arbitrary low temperature, indicating absence of global magnetic
  order}. In the low-temperature phase, the magnetization histogram
becomes nearly constant in temperature and it develops long tails.

\begin{figure}[t!] 
\begin{center} 
 \includegraphics[width=1.\columnwidth]{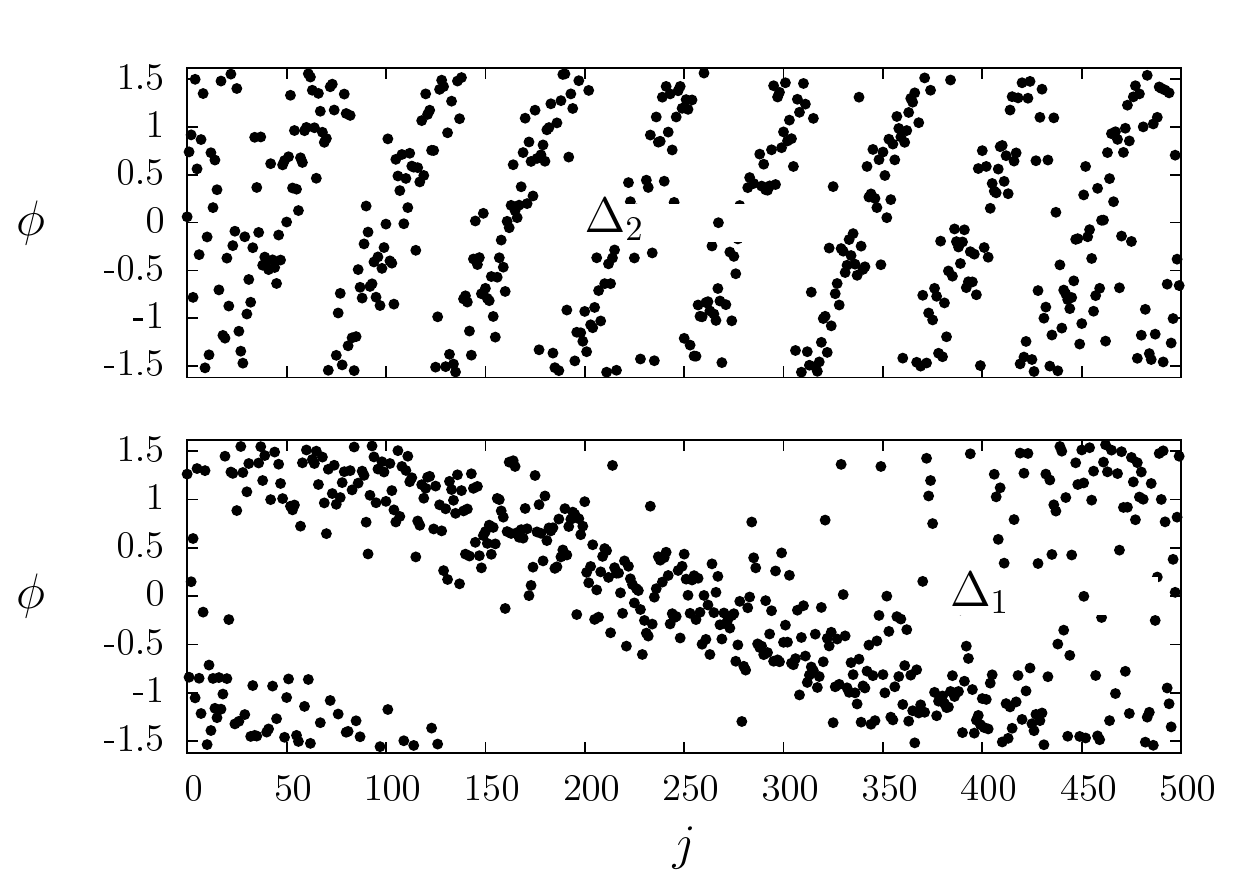}  
\caption{Angular degree of freedom $\phi_j={\rm arg }\, a_j$, versus
  spin index $j$ of two single equilibrated configurations with
  $N=500$ spins in a CT of quadruplets at a common undercritical
  temperature, $T= 0.34$. It is evident the presence of a
  Phase Wave, i.e., the approximated linear dependence of the phase on
  the spin index.  Configurations are shown exhibiting different Phase Wave
  slopes, $\Delta_1$, $\Delta_2$.}
\label{fig:anglesN500}
\end{center}   
\end{figure}

\begin{figure}
\begin{center}
\includegraphics[width=1.\columnwidth]{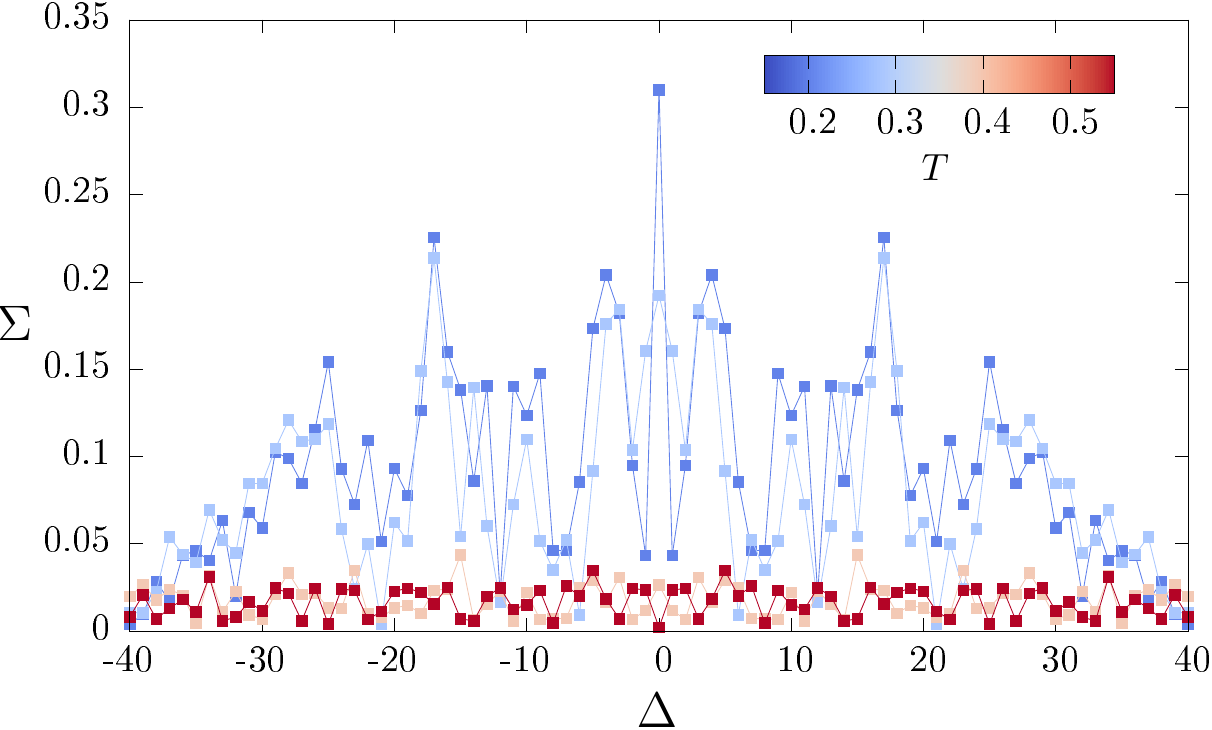}
\caption{The function $\Sigma$, Eq. (\ref{eq:Sigma}), describing the
  proliferation of different Phase Wave slopes, in a CT system with
  $N=80$, $\nq=N^2$, for four temperatures. 
  The different peaks correspond to different
  possible slopes. At supercritical temperatures, $\Sigma$ does not
  present peaks but uncorrelated oscillations, while at undercritical
  temperatures, the position of the peaks is common for all the
  temperatures, and the peak amplitude roughly decreases for
  increasing temperature. }
\label{fig:phiw}
\end{center}
\end{figure}

\subsubsection{Phase Wave and two-point phase correlators}

At the origin of this feature there is a property of the low-$T$ phase
with CT that we call the {\it Phase Wave}. Modes at near-by
frequencies, with small value of $|i-j|$, participate in a larger
number of quadruplets, since the condition Eq. (\ref{eq:myFMC}) is
more frequently satisfied than for distant modes. For this reason,
near-by spins in the frequency are effectively more coupled, and tend to align. This
induces a Phase Wave (in analogy with the ``spin wave'' term in the
context of the $O(2)$ pairwise model): in single low $T$
configurations, the phases of the spins exhibit an approximated linear
dependence with the frequencies $\omega_j$ (or with the spin index,
see Eq. (\ref{eq:freqs}) $\phi(\omega_j) \simeq \phi_0 + \Delta
(\omega_j-\omega_0)$, where $\Delta$ is the phase wave slope, a
configuration-dependent quantity. In Fig. \ref{fig:anglesN500} we
illustrate the phase wave at two different equilibrated
configurations at the same temperature. 

Given a realization of the quadruplet topology, there are some different possible values of the phase
wave slope with a nontrivial probability distribution. We consider, then, the quantity
\beq \Sigma(\Delta) = \left| \sum_{j=1}^N  \< \cos \phi_j \>\, e^{\myi 2\pi j \Delta/N}\right|
\label{eq:Sigma}
.
\eeq
A given value of the slope $\Delta$ is a narrow peak in the function $\Sigma$. 
At a finite temperature, we observe wide peaks in $\Delta$, as a result of thermal fluctuations, at
some privileged values of $\Delta$ depending on the specific realization of
the list of quadruplets (see Fig. \ref{fig:phiw}). Given a realization
of the list of quadruplets, there are peaks at fixed values of
$\Delta$, the amplitude of which increases with decreasing
temperature. Above the critical temperature, on the other hand, the
function $\Sigma$ randomly fluctuates near zero, indicating the lack of
correlation between different spins. 


The Phase Wave is, hence, the microscopic
mechanism for which there is no global $O(2)$ symmetry breaking in the
low-$T$ phase of the CT.

\begin{figure}[t!]                        
\begin{center} 
 \includegraphics[width=1.\columnwidth]{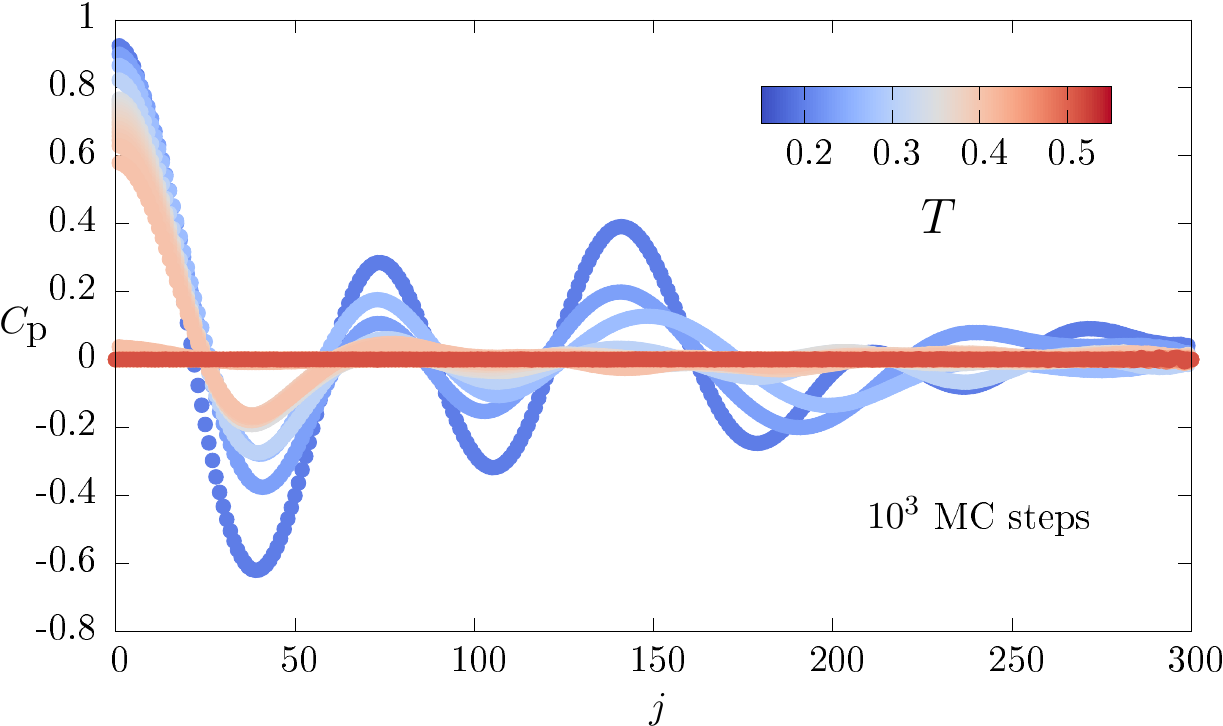}  
\caption{Phase correlation function of a CT system with $N=300$,
  $\nq=N^2$, averaged over small time intervals $\tau=10^3$ MC
  steps. Different curves correspond to different temperatures. 
  While the correlation for supercritical temperatures vanishes up to thermal fluctuations,
  at low temperatures it oscillates with a frequency given by the
  Phase Wave slope. }
\label{fig:Cp}
\end{center}   
\end{figure}

{\it Phase Correlation versus Frequency.} The phase correlation
function helps to further characterize the Phase Wave above
described. Fig. \ref{fig:Cp} reports the phase correlation function
$C_{\rm p}$ for a system with $N=150$ in a CT, for several
temperatures. In the figure, the correlations have been {\it averaged
  over a short number ($\tau \sim 10^3$) of Monte Carlo steps}. While
for $T>T_c$ the phases of different spins are completely uncorrelated,
the correlation is not trivial for $T<T_c$, and presents oscillations
in frequency around zero, in correspondence with the Phase Wave
oscillations: spins near-by in frequency (in spin index) exhibit
strongly correlated orientations, at least in single configurations.  

\begin{figure}
\begin{center} 
 \includegraphics[width=1.\columnwidth]{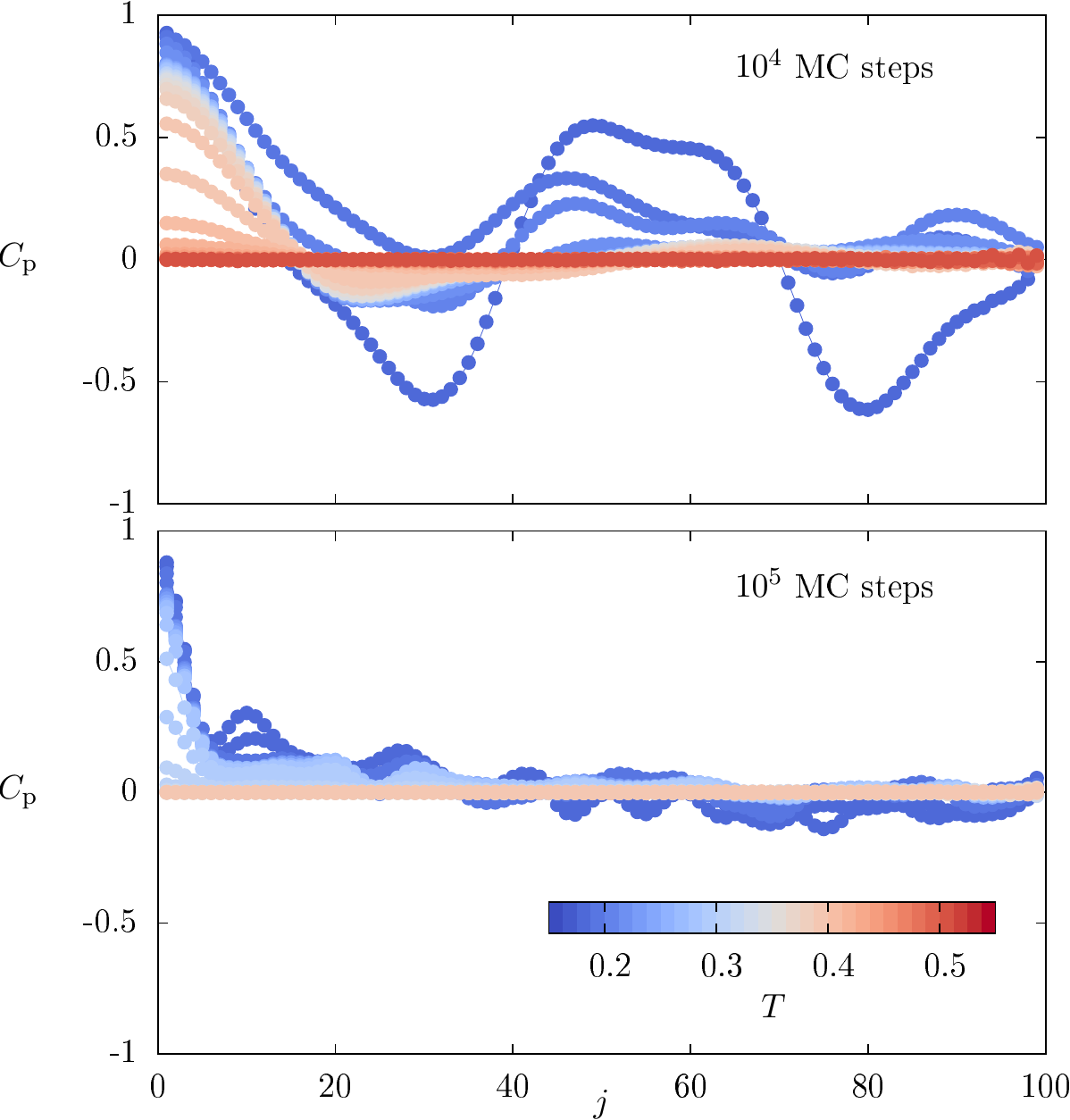}  
\caption{Angular two-point correlators, as in Fig. \ref{fig:Cp}, but
  with $N=100$ and averaged over larger intervals of local-update
  Monte Carlo time ($\tau=10^4$ and $10^5$, upper and lower panel
  respectively).}
\label{fig:Cpintime}
\end{center}   
\end{figure}

The picture, however, turns different when averaging over larger
intervals of time. Our numerical results indicate that in the CT the
sum of the correlations $C_{\rm p}$ over all distances decays to zero
when averaged for arbitrary large Monte Carlo times, at difference
with the HT case (see Fig. \ref{fig:Cpintime} and
Sec. \ref{subsub:dynamics}). This gives strong evidence of the fact
that the two-point angle correlators vanish even at arbitrary low
temperatures. The microscopic origin of this fact is the degeneracy of
Phase Wave configurations with different slopes, so that phase
correlations corresponding to different slopes cancel out.

\subsubsection{Two-point moduli correlations}

\begin{figure}[t!]                        
\begin{center} 
 \includegraphics[width=1.\columnwidth]{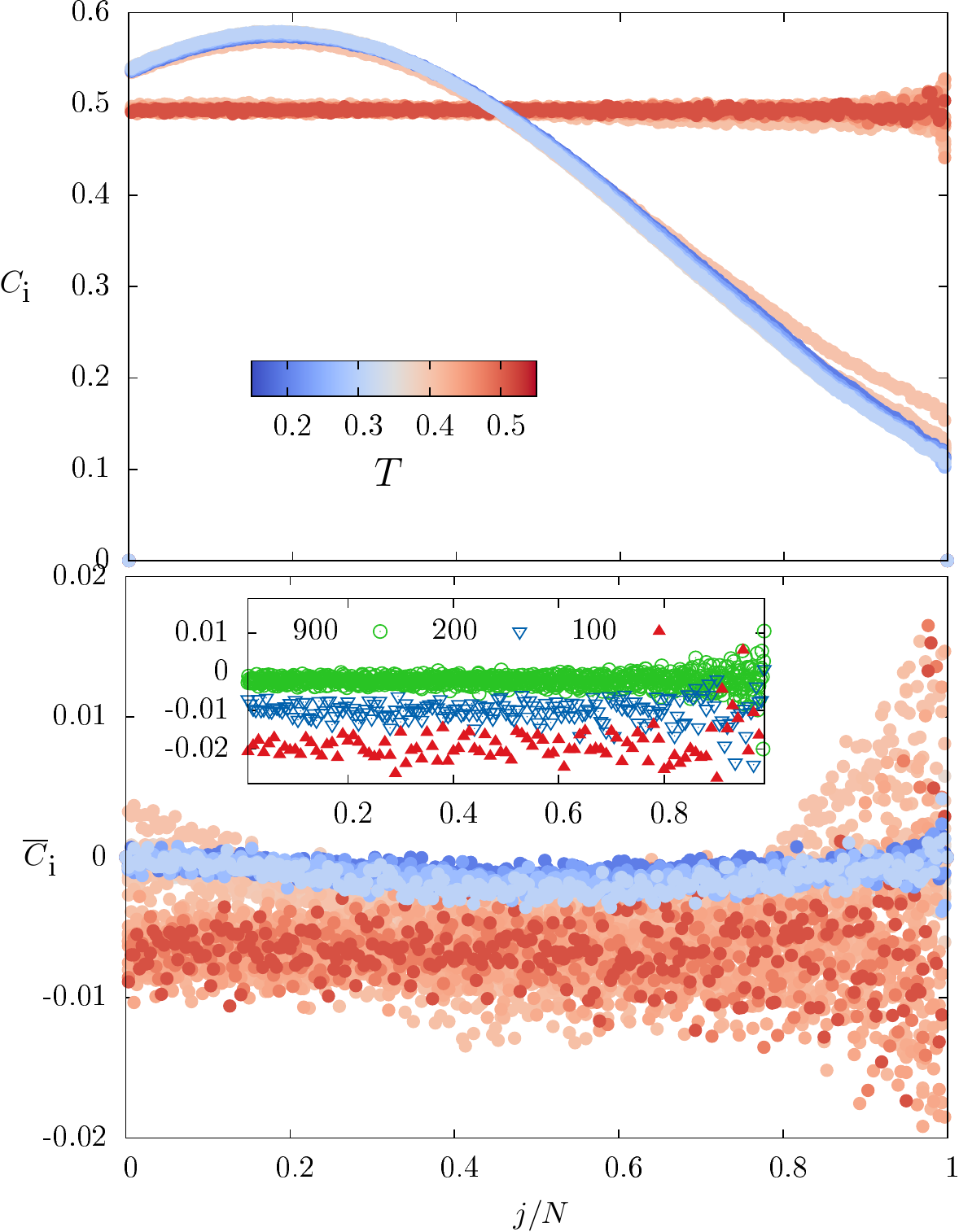}  
\caption{Upper panel: disconnected modulus correlation function, cf., Eq. (\ref{eq:Cs}). $N=300$, $N_4=N^2$, the time averages are over thermalized data in time windows of length $\tau=10^6$.
Lower panel: the connected modulus correlation function (Eq. (\ref{eq:cCs})) for the same system. 
In the inset the connected modulus correlation function at the temperature $T=0.46$ is shown for the sizes $N=900, 200 , 100$,
showing that the function decrease for increasing sizes.
}
\label{fig:Intcorr}
\end{center}   
\end{figure}

As a further insight into the low-temperature phase we present the
behavior of the two-point moduli correlator. As shown in
Fig. \ref{fig:Intcorr}, the disconnected quantity $C_{\rm i}$ is
approximately equal to $(2/\pi)^2$ in the high-$T$ phase, indicating
independence of moduli, while for low temperatures there is a
nontrivial correlation presenting a maximum at a nonzero value of the
spin frequency distance $\omega$, and decaying below the value $(2/\pi)^2$
 for distant spins, which are less coupled and hence less correlated. Since
spins must obey the spherical constraint, the existence of spins with
moduli larger than one implies the
existence of other spins with  moduli less than one. However, the
connected function $\tilde C_{\rm i}$, registering the fluctuations on
top of this general tendency, results to vanish for large system size,
as can be seen for different sizes in Fig. \ref{fig:Intcorr}: for larger and
larger sizes, the values of $\tilde C_{\rm i}$ corresponding to both
phases decrease with the value of $N$, along with the ``gap''
separating the data of both phases, which turns to be a finite-size
effect of the high-temperature phase. 

We conclude that also
moduli-moduli correlators vanish for CT in the thermodynamic limit.

\begin{figure}
\begin{center} 
 \includegraphics[width=1.\columnwidth]{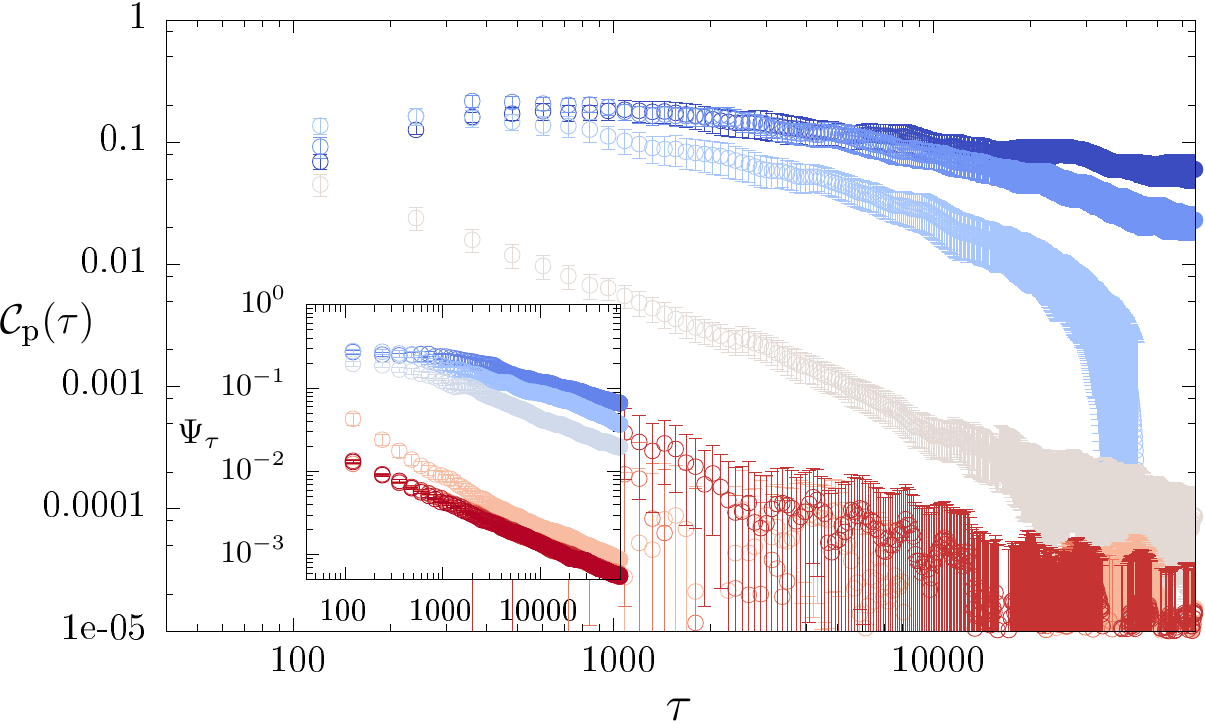}  \\
 \includegraphics[width=1.\columnwidth]{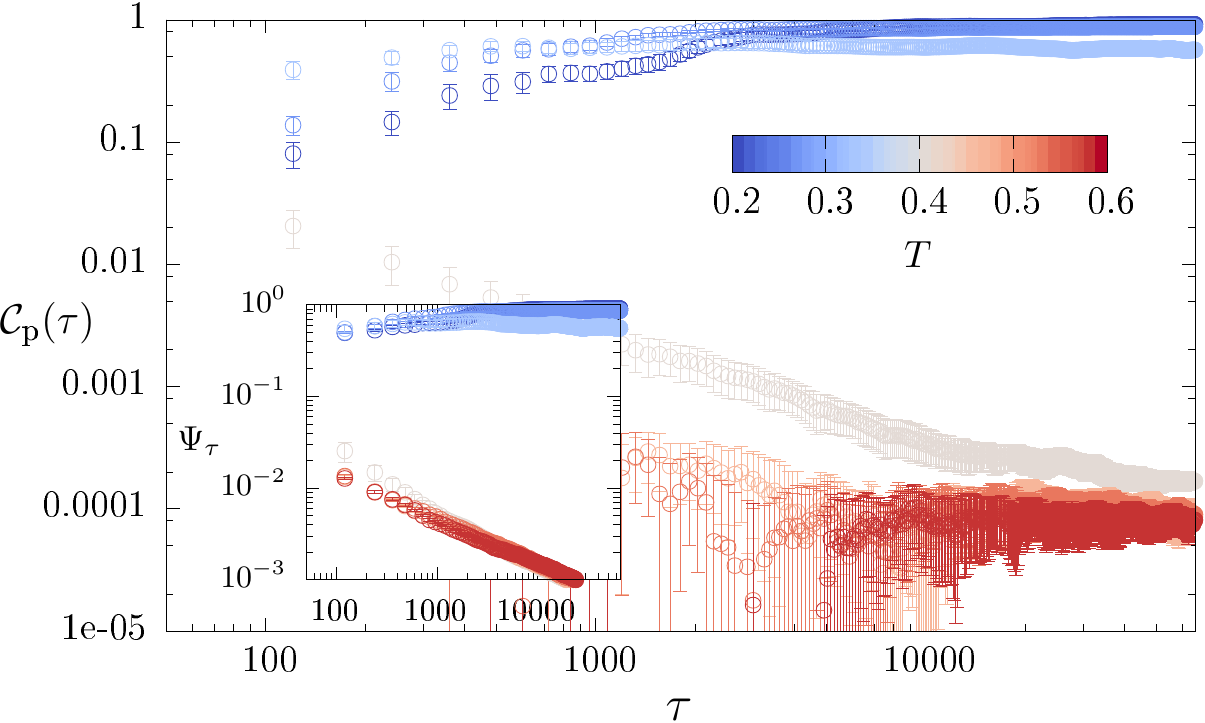}  
\caption{Temporal correlation functions ${\cal C}_{\rm p}$ and $\Psi$ for $N=50$, $\nq=N^2$, CT and HT's (upper and lower panel, respectively), 
for several temperatures (the color code is as in Fig. \ref{fig:maghis}). The finite-size critical temperature is  $T(50)=0.39(7)$. }
\label{fig:corrtime}
\end{center}   
\end{figure}

\subsubsection{Slow dynamics at low temperature}
\label{subsub:dynamics}

We now present numerical evidence of the CT system to exhibit slow dynamics at low temperatures, whose origin is the degeneracy of phase wave configurations with different frequencies. To this aim, we define {\it dynamical} measurements, through the {\it time average}
$\<\cdots\>_\tau=\sum_{\tau'}^\tau(\cdots)/\tau$ over a finite time interval of length $\tau$, in units of local Monte Carlo steps.
For sufficiently large $\tau$, such an average coincides with the thermal average. 
Consequently, we define the $\tau$-Correlation Function for phases and moduli, respectively as
\begin{align}
{\cal C}_{\rm p}(\tau) =& \frac{1}{N}\sum_{r} \Xi_\tau(r) \, , \\
{\cal C}_{\rm i}(\tau) =& \frac{1}{N}\sum_{r} \left[ \langle A_i A_j \rangle_\tau - \<  A_i\>_\tau \< A_j\>_\tau \right]
\, ,
\end{align}
\begin{align}
\Xi_\tau(r) \equiv & \frac{1}{N}\sum_j \Bigl[
\langle \cos(\phi_j-\phi_{j+r})\rangle_\tau 
\\
&
-\langle\cos\phi_j\rangle_\tau
\langle\cos\phi_{j+r}\rangle_\tau
-\langle\sin\phi_j\rangle_\tau
\langle\sin\phi_{j+r}\rangle_\tau
\Bigr] \, ,
\nonumber
\end{align}
 along with the {\it modified $\tau$-long Phase Correlation Function}:
\barray
\Psi(\tau) &=& \sum_{r=1}^N \frac{N}{K(r)}
\, \Bigl|\Xi_\tau(r)\Bigr| 
\earray where $K$ is defined after Eq. (\ref{eq:Cs}). The
functions ${\cal C}_{\rm p,i}(\tau)$ are simply the sum of two-point
correlators in different sites, while $\Psi(\tau)$ is the the sum of
the absolute value of the function $\Xi_{\tau}(r)$ for all the
possible spectral distances $r$. 
Note also that in the limit $\tau \to \infty$ the time average coincides with the 
equilibrium average and, thus, $\Psi(\tau \to \infty)$ is equivalent to $\sum_\omega C_{\rm p} (\omega)$, cf. Eq. (\ref{eq:Cp}).

We stress that the function $\Psi(\tau)$ decays
slower than ${\cal C}_{\rm p}(\tau)$, and it has been defined to
estimate the correlation time of the Phase Wave, as it does not
include the anti-correlation between ``distant'' spins (intrinsic to
the Phase Wave configurations) {\it occurring in single
  configurations}. Both functions, nevertheless, present a
qualitatively similar behavior.

As it has been explained in the previous section, we have found strong
evidence for the thermal average of both $\cal C_{\rm p}$ and $\Psi$
to vanish at low temperature in the CT, and to be nonzero in the
HT. Above $T_c$ they obviously vanish for all topologies, up to
finite-size effects. This is illustrated in Fig. \ref{fig:corrtime}
for a $N=50$ system in the CT case, where correlations decay towards
zero for sufficiently large times.

A remarkable feature of the temporal correlation functions is that, at
least for CT, both $\cal C_{\rm p}$ and $\Psi$ decay slower and slower
as temperature decreases. This is also reflected in the probability
distribution of the Phase Wave correlation time, $\tau_\phi$, defined
as the time employed by $\Psi(\tau)$ to decay below a given
threshold. Such a distribution develops long tails as temperature
decreases, as shown in Ref.  \onlinecite{Antenucci2014statistical}.

 An explanation for such a behavior is provided by the dynamical
 measure of the function $\Sigma$. Its estimation in equilibrium dynamical
 simulations over a time window such that $\cal C_{\rm p}$ in
 Fig. \ref{fig:corrtime} has not yet decayed, {\it presents just few
   peaks} or even a single peak only, corresponding to the few
 different Phase Wave configurations with fixed slope in which the
 system remains trapped during few thousands of local MC steps. 
 In this situation the use of a nonlocal update, as the Parallel Tempering
 algorithm, is essential to thermalize the system (to get it decorrelated)
 in a feasible number of MC steps ($\sim 10^4$ for a system with $N=50$), 
 recovering the multiplicity of peaks in Fig. \ref{fig:phiw}.
%
These facts suggest a dynamical picture of the CT
system according to which, at low temperatures, different Phase Wave
slopes are degenerated and correspond in some way to different minima
in the potential energy landscape, so that the time to escape from one
of them dramatically increases with decreasing $T$. 

A careful sight suggests that a slow dynamics may be present
also in the HT case, whose origin is, however, different, 
being towards a nonzero value for the correlation. 
The analysis in the HT case is more difficult since it
requires the knowledge of the thermalized probability distributions of
${\cal C}_{\rm p}$, $\Psi$ at different temperatures.  A deeper study
is necessary to describe the dynamics of both cases in an accurate way.

The moduli temporal correlation function ${\cal C}_{\rm i}$  presents but quite short relaxation times even at low temperature in both HT and CT, 
indicating that the moduli dynamics is irrelevant in the emergence of large timescales.

\begin{figure}
\begin{center} 
 \includegraphics[width=1.\columnwidth]{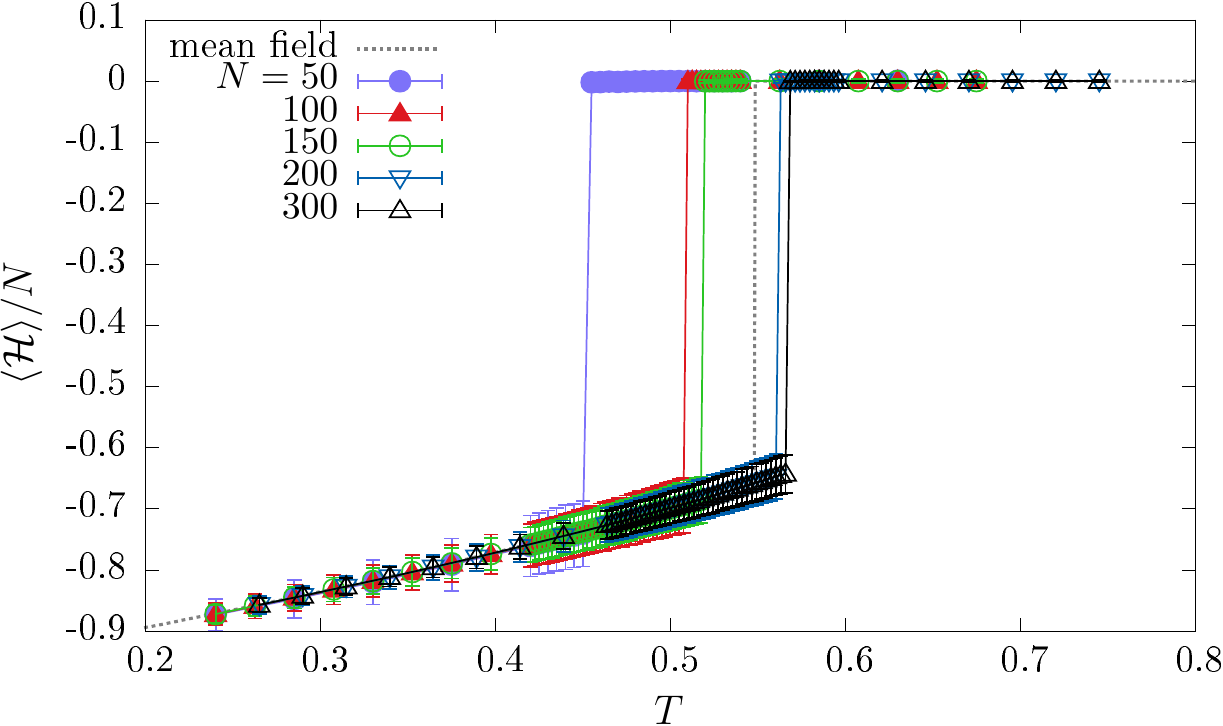}  
\caption{Energy of the XY model in a dense, homogeneous set of quadruplets ($\nq = N^2$) for several sizes. The two largest have a critical transition temperature larger than the mean field prediction. 
The data are indistinguishable  from the mean field result for $T<T_c^{\rm mf}$.}
\label{fig:XYdense}
\end{center}   
\end{figure}

\section{Numerical results for the XY model}
\label{sec:resultsXY}

The XY model with four-body interactions (defined by the Hamiltonian Eq. (\ref{eq:HXY}) with quenched amplitudes, $A_j=1$ for all $j$), 
presents as well a rich and interesting phenomenology, that we now resume. 

{\it Dense Homogeneous Topology.} As we have explained in the previous
section, the moduli dynamics at low temperatures does not play
any essential role in the thermodynamics of the Spherical Model for a
{\it dense} ($\nq \sim \mathcal{O} (N^{\geq 2})$) set of quadruplets: in the
low-temperature phase the moduli are more and more homogeneous and
equal to one at lower and lower temperatures. The behavior of the
$p=4$ XY model in dense HT's, as one could expect from this argument,
is indeed qualitatively identical to that of the Spherical Model:
there is a discontinuous phase transition separating a phase with
uncorrelated angles, and a low-$T$ magnetized phase with $O(2)$
symmetry breaking. The finite-size critical point $T_c(N)$ is 
obviously higher than the Complex Spherical Model case (see Fig. \ref{fig:XYdense}). 
In the dilute (though dense) version, $\nq \sim \mathcal{O} (N^2)$, we have observed how the mean field solution accurately reproduce  
the numerical results for energy and magnetization, with the exception of the transition temperature, 
which may be higher than the mean field value (see sizes $N=200$ and $300$ in Fig. \ref{fig:XYdense}). 
In the fully connected case the critical temperature is compatible with the mean field value.

\begin{figure}[t!]                        
\begin{center} 
 \includegraphics[width=1.\columnwidth]{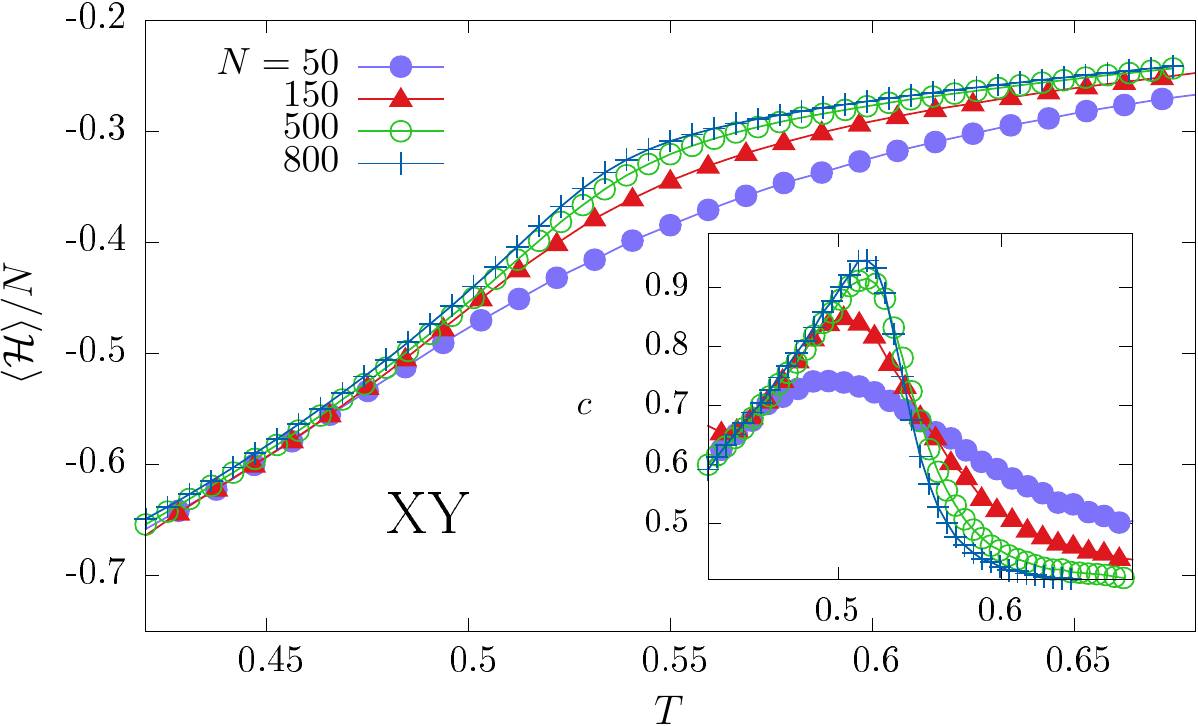}  
\caption{Energy of the XY model in sparse ($\nq= N$), homogeneous sets of quadruplets, for four sizes. 
The energy is continuous at the transition.
Inset: Specific Heat for the same systems.  }
\label{fig:XYsparse}
\end{center}   
\end{figure}

\begin{figure}[t!]                        
\begin{center} 
  \includegraphics[width=1.\columnwidth]{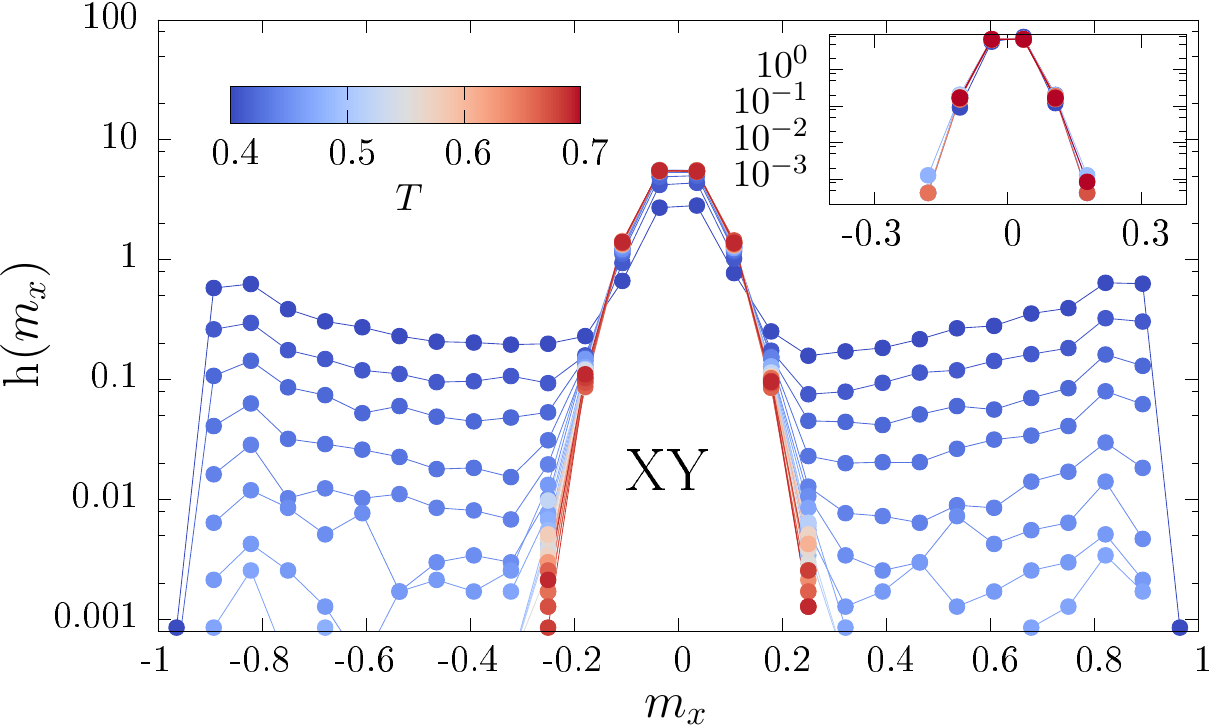}  
\caption{$m_x$ histogram of the XY model in a sparse ($\nq = N$), homogeneous set of quadruplets, for $N=150$. 
Inset: $m_x$ histogram for the size $N=500$.}
\label{fig:XYsparsehis}
\end{center}   
\end{figure}

{\it Sparse Homogeneous Topology.} 
We have also considered the case with high dilution, so that the number of quadruplets is $\nq\sim \mathcal{O}(N^{< 2})$.
For the Complex Spherical Model, one obtains a non-equipartite condensation in such a topology, as
explained in Sec. \ref{sec:threshold}.
In the XY case, our simulations provide instead evidence that the system exhibits the mean-field behavior for $\nq\sim \mathcal{O}(N^{>1})$.
In the extensive, \emph{homogeneously sparse}, case $\nq\sim \mathcal{O}(N)$, instead, we observe evidence for the onset of  a second-order phase transition,
separating two unmagnetized phases. Remarkably, the effect of diluting, until
reaching sparseness, has the effect of preventing the symmetry breaking. 
The energy presents no discontinuity while the specific heat is an increasing function of $N$ at the transition, cf. Fig. \ref{fig:XYsparse}. 
The magnetization histograms reveal absence of angular symmetry breaking, with long tails that appear continuously at low temperatures
and whose magnitude decreases with the size of the system, cf. Fig. \ref{fig:XYsparsehis}.
The resemblance of the \emph{sparse} case with the unbroken symmetry in the pairwise XY model in two dimensions is discussed in the next section.

\begin{figure}[t!]                        
\begin{center} 
 \includegraphics[width=1.\columnwidth]{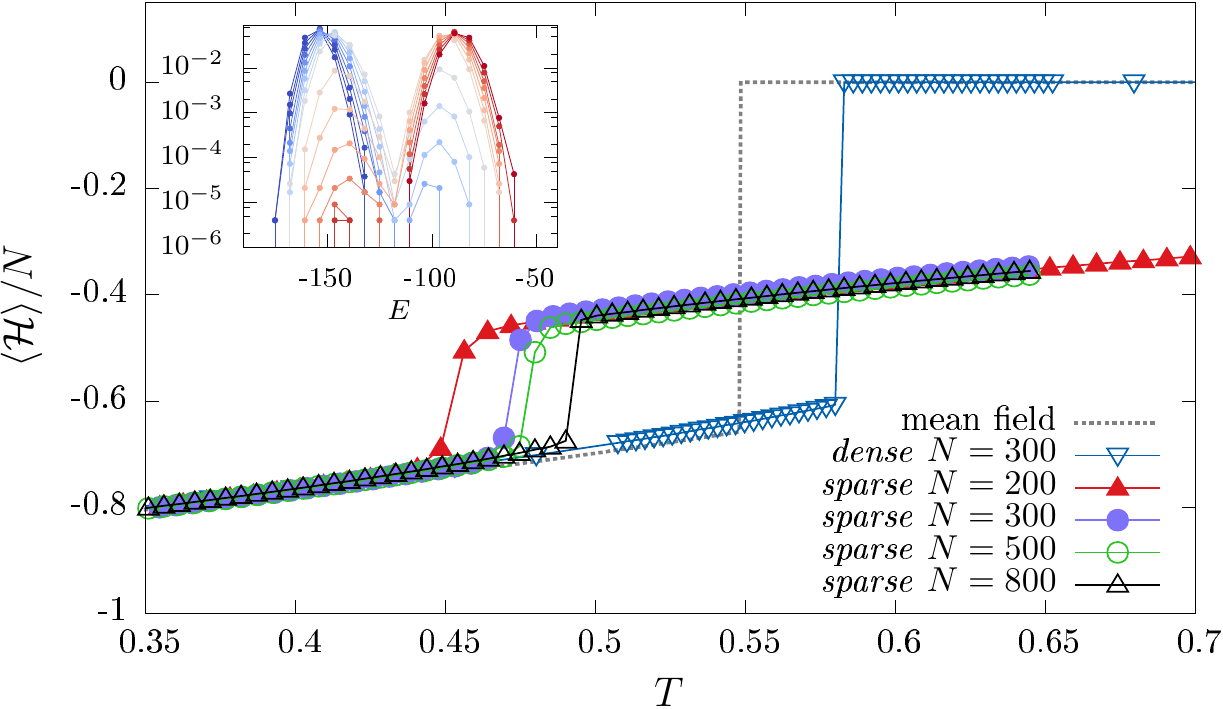}  
\caption{XY model energy in CTs for several sizes and two types of
  dilution, dense ($N_4 = N^2$) and sparse ($N_4=N$). The inset shows the energy histogram in
  the sparse case with $N=200$ and several temperatures in the range
  $[0.38:0.51]$. In the presence of CT the transition remains first
  order even in the sparse system, with the energy in the low temperature phase
  coinciding with the mean field theory (see main text).}
\label{fig:XYCT}
\end{center}   
\end{figure}

{\it Correlated Topology.} 
Remarkably, for both $\nq \sim \order (N)$ and
$\nq \sim \mathcal{O}( N^{\geq 2})$, our numerical analysis suggests that the phase
transition remains discontinuous, with a low-temperature phase
characterized by the absence of magnetization and the presence of phase waves, as in the Complex Spherical Model case in a dense CT. 
In the presence of a CT, our results indicate that the sparseness of the list  of quadruplets (i.e., $\nq \sim \order (N)$ in this case)
only changes the approach to the critical point from \emph{high temperatures}, 
where we observe an ordering with negative energy different from the phase wave,
while the transition remains first order.
We present the finite size $E(T)$ curves in Fig. \ref{fig:XYCT}. 
From our data, we have
concluded that the transition remains first order in the sparse case,
since there is no finite size indication of divergence in the susceptibility $\chi = N ( \<|m|^2\>-|\<m \>|^2 )$, and
since the energy histogram presents two separated peaks, with a
coexistence region, as can be seen in the inset of
Fig. \ref{fig:XYCT}. The magnetization histograms are qualitatively
identical to those of the Complex Spherical Model in CTs, in
Fig. \ref{fig:maghis}.

\vspace{0.5cm}

The whole picture on the type of low-temperature
behavior and the symmetry conservation for all of our models happens
to be rich and unexpected, and it is outlined in Table
\ref{table:transitions}.

\begin{table}[t!]
\begin{tabular}{||c|c|c|c|c||}
\hline
\hline
model & topology & $\nq$ & transition & $m(T<T_c)$ \\
\hline
CSM & HT & $\order (N^{\geq 2})$ & $1^{st}$-order & $\ne 0$ \\
CSM & HT & $\order (N^{< 2})$ & non-Eq. cond. & - \\
CSM & CT & $\order (N^{\geq 2})$ & $1^{st}$-order & $=0$ \\
CSM & CT & $\order (N^{< 2})$ & non-Eq. cond. & - \\
\hline
XY & HT & $\order (N^{> 1})$ & $1^{st}$-order & $\ne 0$ \\
XY & HT & $\order (N)$ & $2^{nd}$-order & $=0$ \\
XY & CT & $\order (N^{> 1})$ & $1^{st}$-order & $=0$ \\
XY & CT & $\order (N)$& $1^{st}$-order & $=0$ \\
\hline
\hline
\end{tabular}
\caption{
Nature of the transitions and of the correlators in the low-$T$ phase for the different considered models, as emerges from the numerical analysis. 
Whenever $m=0$, also the two-point correlators ${\cal C}^{{\rm i},{\rm p}}$ vanish. }
\label{table:transitions}
\end{table}

A remarkable fact of the results of
Fig. \ref{fig:XYCT} is that the energy of the dense case coincides with
the mean field energy in the whole range $T< {\rm min}\{T_c(N),T_c^{\rm
  mf}\}$, although the finite-size transition temperature can be
larger than the mean field solution. Such an agreement between the CT
and the HT observables (and, incidentally, between both and the mean field
theory) was absent in the Complex Spherical Model (see the precedent section),
indicating that the differences between CT and HT is attributable to
the moduli dynamics. What is more, we observe that also in the sparse
case there is an agreement between mean field theory, dense CT and
dense HT for sufficiently low temperature. 


\section{Analogy with the Abelian lattice gauge theory}
\label{sec:analogy}

In the introduction we have mentioned the fact that the
three-dimensional Abelian lattice gauge theory presents a second-order
phase transition, mappable to the 2D Kosterlitz-Thouless
transition. The low-temperature phase is unmagnetized, a property
which follows from the model gauge invariance via Elitzur's theorem,
which states that non-invariant observables under a gauge
transformation present vanishing expected value in a gauge-variant
system. We believe this mechanism to be the origin of the vanishing of
the magnetization also in our 4-XY model in a homogeneous sparse topology,
mentioned in the previous section. According to this argument, the
stochastic set of homogeneous quadruplets acquires a kind of gauge
invariance under some type of transformations. For example, it is
possible that in a sparse list of random quadruplets there is a
proliferation of sets of four spins which, although not forming a
quadruplet, occupy the bonds of four neighboring quadruplets (as the
sets of four spins on which the lattice gauge transformations act). These arguments justify the fact that in presence of
topological correlations, the 1-point (magnetization) and the 2-point
(phase and intensity correlators) operators vanish, since they are not
invariant under gauge symmetry transformation, involving four spins. On the other hand, four-point correlators, as the different terms in the Hamiltonian, are nonzero in general.

In any case, we stress that such a symmetry does not completely forbid
the presence of magnetized configurations: in
Fig. \ref{fig:XYsparsehis} one observes two maxima of the distribution
$h(m_x)$ at nonzero values of $m_x$. These magnetizations, however,
are much less probable than the most probable value at $m_x=0$.  An
analogous mechanism could be behind the vanishing of the average
magnetization found in both XY and Spherical models in the presence of
topological correlations. In this case, the transformations leaving
the total energy invariant (up to fluctuations) would depend on the
frequencies, and would be no longer local but global transformations
connecting Phase Wave configurations with different allowed slopes.

\section{Connection with optics and possible experimental realizations}
\label{sec:optics}

Interpreted from the point of view of optics, the results of our
analysis lead to several straightforward consequences in the field of
multimode laser formation.  Perhaps the most immediate result, not
captured by approaches that neglect the role of the frequencies,
is the existence of a correlated phase without global $O(2)$ order,
whose microscopic origin is the Phase Wave.  
We now explain how this
novel phase can have experimentally accessible consequences in the
form of a phase delay of the ultra-short electromagnetic pulses
resulting from the nontrivial mode-locking in the presence of
FMC.\cite{Antenucci2014statistical} Such a temporal delay should be
experimentally accessible, {as similar carrier phase delays are
measured even in ultra-short lasers.\cite{Barbec2000} }

\subsection{Phase Delay and Phase Wave}

Let $\tau$ be the time measured in units of the time interval between
two pulses, which in the statistical physical framework can be
associated to a microscopic unit of time evolution, for example a
 Monte Carlo step. Let $a_n(\tau)=A_n(\tau)e^{\myi
  \phi_n(\tau)}$ be the $n$-th electromagnetic mode at the Monte Carlo
time $\tau$. Consider also the microscopic time unit $t\ll \tau$
describing the evolution of the electromagnetic pulse,
whose form is:
\begin{equation}
 E(t|\tau) =\sum_{n=1}^N A_n(\tau) e^{\imath [2\pi \omega_n t +
  \phi_n(\tau)]} \, .
\label{eq:pulse}
\end{equation} 

\begin{figure}[t!]
\begin{center}
\includegraphics[width=1.\columnwidth]{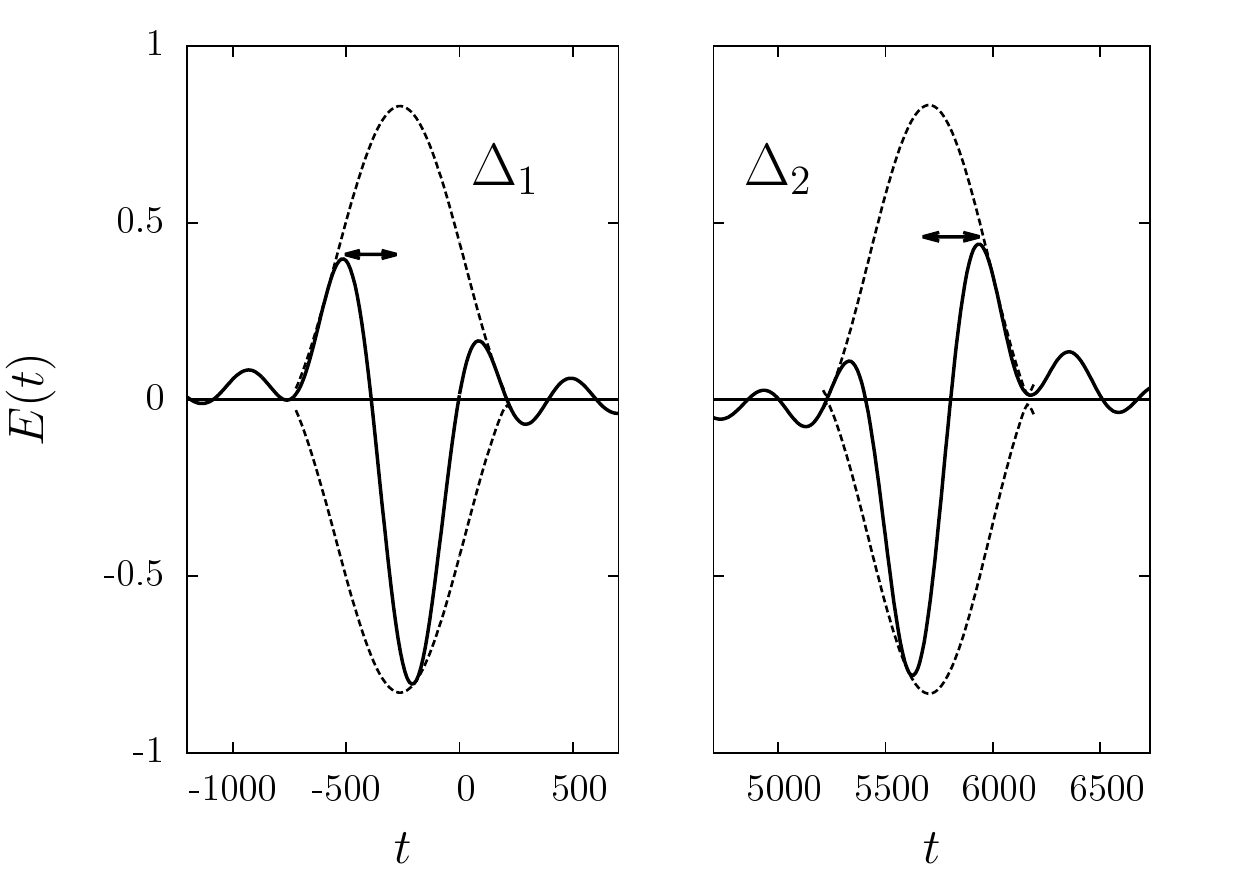}
\caption{Electromagnetic pulse in time, Eq. (\ref{eq:pulse}), with
  $\omega_0=374$, $\delta=1$ (cf. eq (\ref{eq:freqs})), in correspondence of the two Phase Wave
  configurations in figure \ref{fig:anglesN500}, with slopes
  $\Delta_1$ and $\Delta_2$. The carrier-envelope delay indicated as horizontal
  arrows is a function of $\Delta$ and $\omega_0$.}
\label{fig:pulse}
\end{center}
\end{figure}

In the low-temperature phase (i.e., the mode-locking phase at high
pumping rate) of a system with HT, all electromagnetic modes exhibit a
common phase $\phi_n=\phi$ up to thermal fluctuations, and there is no
phase delay in the resulting $E$. On the other hand, the non-trivial
ML induced by the CT is such that the phase velocity $\di E/\di
t|_{t_0}$ changes from pulse to pulse, where $t_0$ is a reference time
with respect to the position of the maximum envelope at a given
$\tau$. The time delay of the field with respect to the envelope is a
nontrivial function of the Phase Wave slope $\Delta$ and of the
central frequency $\omega_0$ (see Eq. (\ref{eq:freqs})). We show in
Fig. \ref{fig:pulse} the form of the pulses at different thermalized
configurations characterized by different $\tau$'s, and their
corresponding phase waves, from which the fields $E$ have been
calculated through Eqs. (\ref{eq:freqs},\ref{eq:pulse}).

In summary, the relaxation of the narrow band approximation requires
the introduction of the role of mode frequencies, through the FMC,
Eq. (\ref{eq:FMC}). We have seen in Secs. \ref{sec:resultsSM},
\ref{sec:resultsXY} how this, in turn, induces the phase wave
mechanism. We propose that, whenever the role of the frequencies of a
multimode laser is not negligible, and if the present model
effectively describes the pulse formation (as it is the case of the
passive mode-locking laser in a closed cavity, which satisfies these
two conditions), it should be observable a carrier-envelope delay of
stochastic nature, of a magnitude changing, in general, from pulse to
pulse (as in Fig. \ref{fig:pulse}). 
Such single pulses dynamics and also its relationship to experimental measurements
of the average signal over several (thousands) pulses, is currently under investigation.

\subsection{Non-equipartite condensation}

The non-equipartite condensation phenomena may manifest in
experimental circumstances, more complicated than the multimode cavity
resonant, such that the dilution of interaction between modes can be
tuned through some mechanism. In a random laser, this is determined by
the spatial separation between electromagnetic modes, since the
coupling between four of them is proportional to their spatial
overlap.\cite{Conti2011}  In a situation in which the leading interaction is
given by the disordered version of Eq. (\ref{eq:H4new}), one expects to observe, 
by varying the spatial concentration of modes, an {\it abrupt transition} from a regime with single isolated peak spectra, with a few number
of very intense modes, to  a  continuous spectra in which the optical intensity is roughly equidistributed among different modes.

In this spirit, we propose an interpretation of the results of the
experiment performed in Ref. \onlinecite{Leonetti2011},  the
first experimental observation of the onset of mode-locking order in
random lasers. In this experiment, a sample of nanoparticles is
immersed in a gain medium, and the pumping protocol is such that the
spatial region of the sample to be pumped can be continuously
enlarged, though maintaining the overall optical power constant.  In
this way, when a large fraction of the sample is illuminated, the
onset of a continuous collective spectra is observed, corresponding to
a large amount of overlapping modes.  When, instead, only part of the
sample is pumped, the activated modes are low-overlapping in space,
their interaction is sparse, and the intensity behavior is as that of the non-equipartited phase described in Sec. \ref{sec:threshold}.

\subsection{Gain and Intensity Spectrum}
\label{sec:gain}

One of the most easily accessible experimental quantities in laser setups is the
intensity spectrum of the signal, $I(\omega)$.  In our framework the
spectra can be directly evaluated so to allow for a straightforward comparison. 

For the study of the spectra, it is interesting to
consider the introduction of a non-flat {\it gain
  curve},\cite{HausPaper} which generalizes the Hamiltonian Eq. (\ref{eq:H4new}) in the
following way:
\begin{align}
 {\cal H} = - \sum_s G_s A_s^2  -  \frac{N}{8\nq} \sum_{spqr}\ {\cal A}_{s p q r}  \, A_{s} A_{p} A_{q} A_{r} \nonumber  \\ 
 \cos (\phi_{s}-\phi_{p}+\phi_{q}-\phi_{r} )
\label{eq:Hgain}
.
\end{align}

We consider Gaussian gain curves $G_s \equiv G(\omega_s)$, $G$ being a
Gaussian distribution with the maximum at the center $\omega_0$ of the
spectrum, and variance $\sigma_g$. In experiments, the temperature is
typically constant, while the optical energy $\epsilon$ is ranged. To
correctly compare with our simulations, where $T$ varies at constant
$\epsilon$, we measure the intensity spectrum as $I(\omega_j) = \<|a_j|^2\>
/ \sqrt{T} $.  In this case, to be consistent with the photonic
counterpart, one also has to consider a temperature rescaled
gain curve: $G(\omega,T) = T G_0 (\omega) $, with a reference gain
curve $G_0 (\omega)$.

We now summarize the results of our numerical analysis of the
Hamiltonian Eq. (\ref{eq:Hgain}). As a first observation we point out
that the system behavior is robust against the inclusion of the local
gain term: the critical properties and the general features of
thermodynamic phases described in the previous sections remain
unchanged.

\begin{figure}[t!]                        
\begin{center} 
\includegraphics[width=1.03\columnwidth]{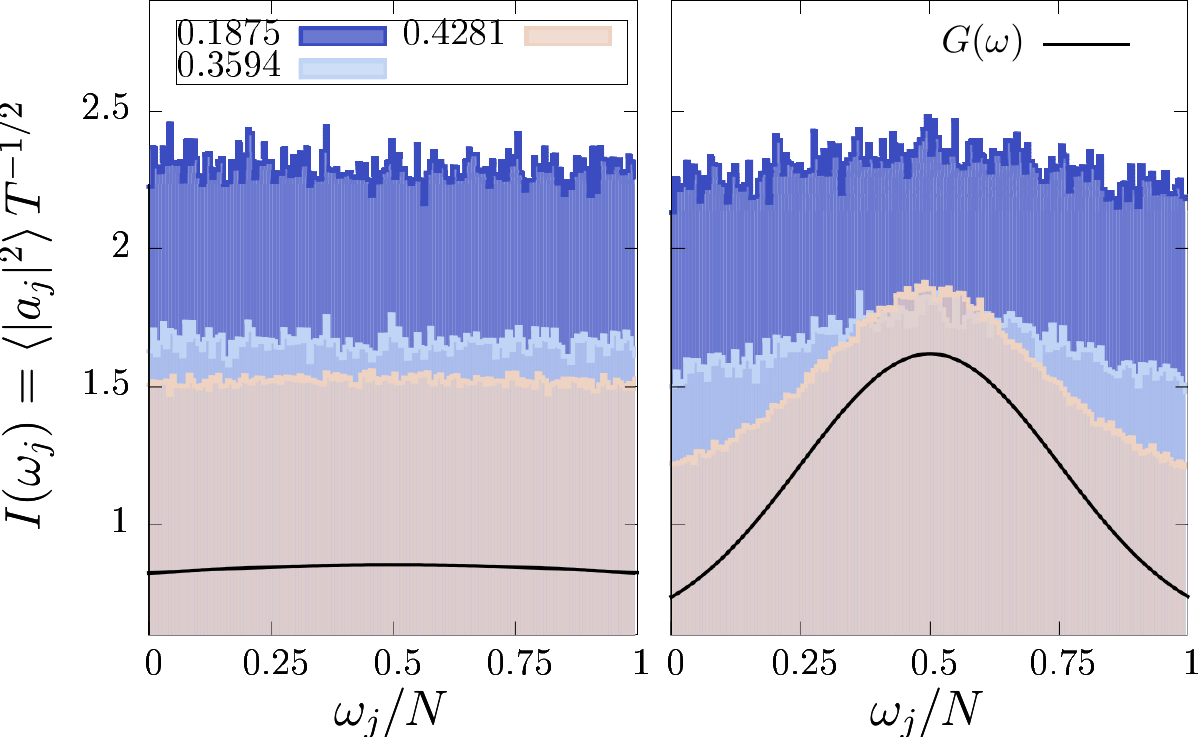}  
\caption{
  Intensity spectra for HT system with $N=150$ and $N_4 = N^2$ at three different temperatures.  For this system
  the transition is at $T_c (150) = 0.369(3)$.  The gain curve is
  Gaussian with mean in the center of the considered frequencies.
  Left: Gain profile with larger variance, $\sigma_g= N$.  Right: Gain
  profile with smaller variance, $\sigma_g = N/4$.  }
\label{fig:spectra_HT}
\end{center}   
\end{figure}

\begin{figure}[t!]                        
\begin{center} 
\includegraphics[width=1.03\columnwidth]{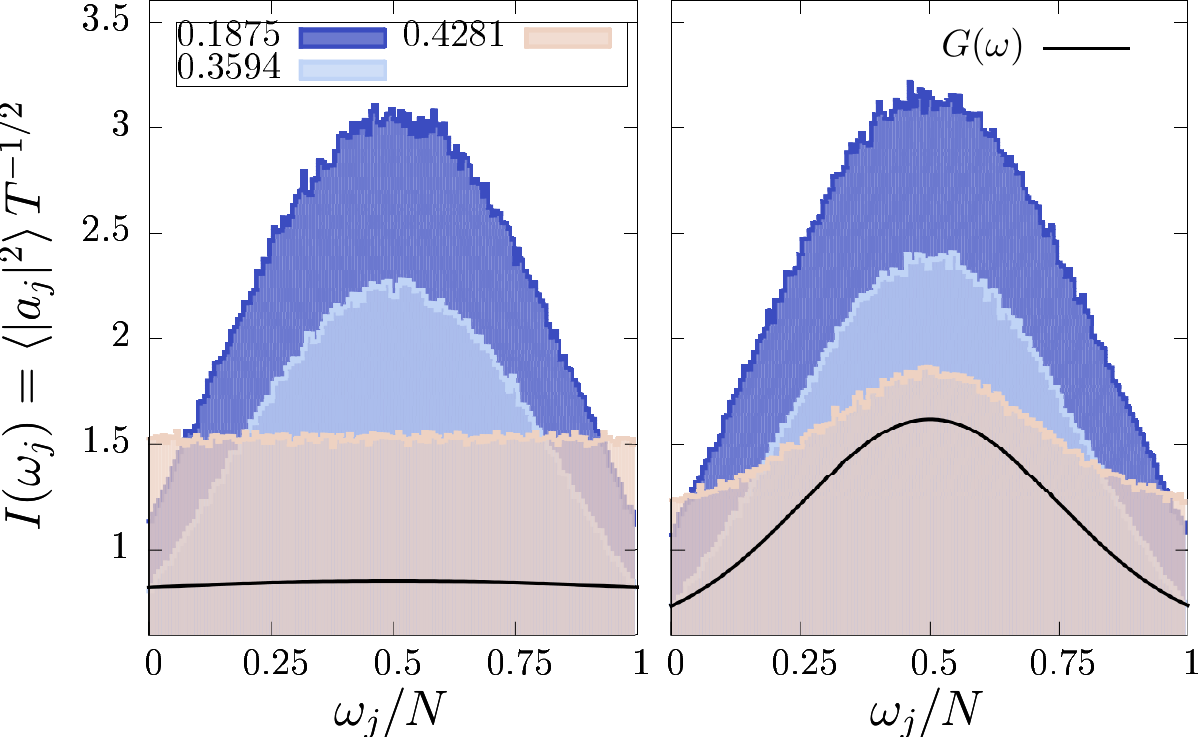}  
\caption{
  Intensity spectra for CT system with $N=150$ and $N_4 = N^2$ at three different temperatures.  For this system
  the transition is at $T_c (150) = 0.386(3)$.  The gain curve is
  Gaussian with mean in the center of the considered frequencies.
  Left: Gain profile with larger variance, $\sigma_g = N$.  Right:
  Gain profile with smaller variance, $\sigma_g = N/4$.  }
\label{fig:spectra}
\end{center}   
\end{figure}

In the IW regime the intensity spectrum is rather influenced by the
shape of the gain curve, see Figs \ref{fig:spectra_HT} and
\ref{fig:spectra}.  In general, the transition causes an abrupt change
in the intensity spectrum.  Above the lasing threshold, in the ML regime, the
intensity spectrum is mainly determined by the topology of the
interactions and it is stable against the introduction of a non-flat
gain curve.  For HT, the intensity spectrum is flat for high enough
pumping, see Fig \ref{fig:spectra_HT}. This reflects the fact that in
HT the frequencies do not play any role, besides the gain curve, and
this role becomes no longer dominant in the ML phase. In particular,
comparing to the case of  an approximately flat gain curve, the spectrum does not change 
above the transition threshold   (cf. left panel of Fig \ref{fig:spectra_HT}).

\begin{figure}[t!]                        
\begin{center} 
\includegraphics[width=0.85\columnwidth]{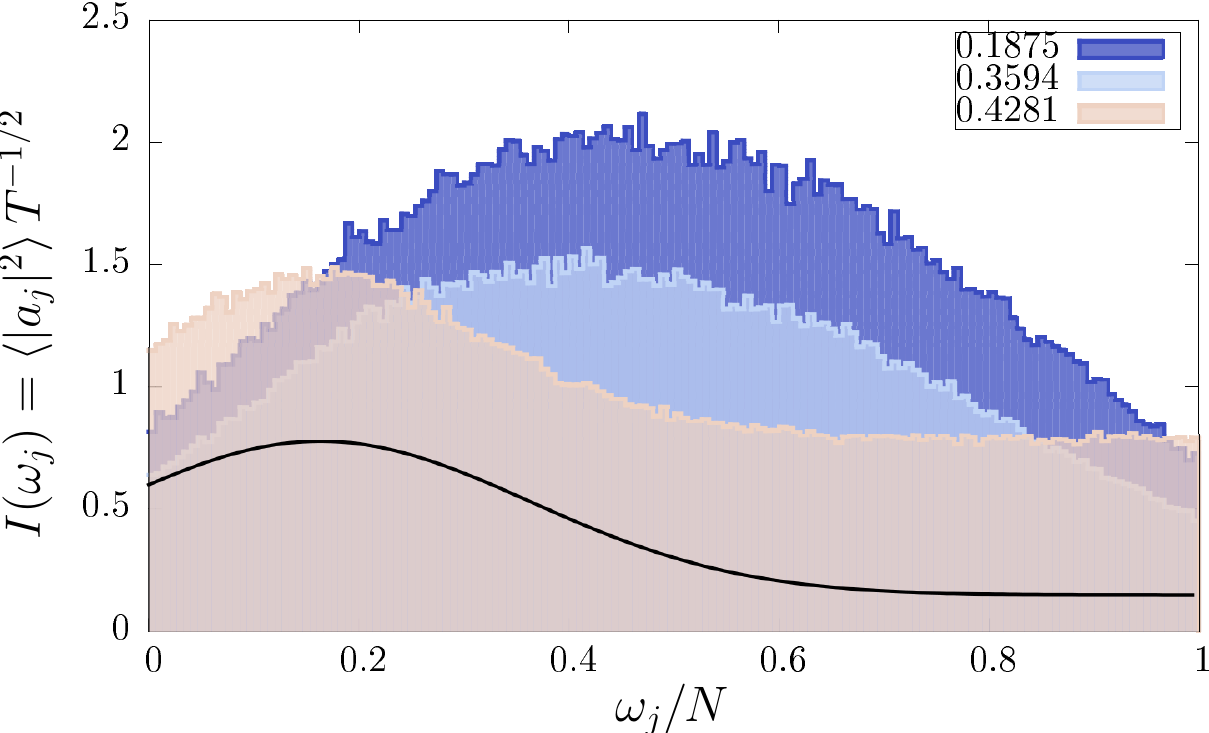}  
\caption{
  Intensity spectra for the same system of Fig. \ref{fig:spectra} but
  with a a gain curve $\sigma_g = N/4$, centered in $\omega_j= N/6$.  }
\label{fig:spectraNonCentered}
\end{center}   
\end{figure}

The intensity spectrum has full sense in the CT, where, instead, the
frequencies play a relevant role in determining the topology.  
In this case, the transition is generally more abrupt in the intensity spectrum.
Above the threshold the spectrum is peaked around the central
frequencies disregarding the shape of the gain (Fig. \ref{fig:spectra}), as the
modes at the central frequencies are effectively strongly coupled
(cf. Fig \ref{fig:quadruplets}).  In other words, the ML spectrum
shape observed in experiments results from our analysis to be a direct
consequence of the frequency-dependent mode interactions resulting
from the FMC.  In Fig. \ref{fig:spectraNonCentered} this outcome is
emphasized considering a gain curve with an average
different from the central frequency of the amplified spectrum: both 
the frequency of maximum intensity and the whole
shape of the spectrum  abruptly
change at the ML threshold.

The observed effect may furnish a theoretical mechanism to
explain the so-called {\it gain narrowing}
phenomena.\cite{HausPaper,Horowitz1994,Horowitz1994linewidth} 

\subsection{Other possible experimental consequences}

In this section we propose two further aspects of our analysis that
may have a direct experimental consequence.  The first one is the possibility
of experimentally measure the { vanishing of two-point (phase and
  intensity) correlators}.  If the correlation measurements are
averaged over times much larger than a light round-trip in the cavity,
the vanishing of two point correlators should be observable.  The
vanishing of two-point correlators may, then, sign the dominance of
the nonlinear interaction mediated by the FMC, which leads to null
two-point functions, as emerges from our analysis in
Sec. \ref{sec:resultsSM}.  Even in random lasers, intensity-intensity
correlations can be measured (see, for example,
Refs. \onlinecite{ghofraniha2014experimental,Leonetti2013,Wu2007,Mujumdar2007}).
On the other hand, phase-phase correlations are measured in
conventional lasers with standard techniques\cite{trebino2000frequency} and, in principle, the
Phase Wave could be observed through phase correlation oscillations
(as in Fig. \ref{fig:Cp}) if a sufficiently high time resolution is
achievable.

Secondly, according to the analysis presented in
Sec. \ref{sec:resultsSM},  the metastable phase in the ML regime is
expected to decrease with the size of the system.
This is observed even in the HT, where the role of frequencies is irrelevant. 
In the optical counterpart this would imply that the region of Optical Bistability \cite{gibbs1985optical} 
should decrease as the number of modes in the multimode mode locking setup increases.  
The other way round, the dependence/independence of this phenomena on the number of
nodes could be used to infer whether the Optical Bistability is a
consequence of the (finite-size) metastability observed in our
simulations, or whether its origin is different.

\section{Conclusions and perspectives}
\label{sec:conclusions}

To the best of our knowledge, the present analysis is the first study
of vector statistical models with four-body interactions beyond the
mean field approximation. From our numerical study it emerges that
these systems present a very rich phenomenology, among
which we highlight: the absence of global symmetry breaking in the
presence of quadruplet correlations; 
the absence of symmetry breaking and the
smoothness of the transition in the XY model for $\nq \sim \order(N)$ homogeneous interacting
terms; the non-equipartite condensation of the Complex Spherical model
on sparse ($\nq \sim \order(N^{<2})$) graphs; the slowing down of the dynamics in the
low-temperature phase. Other rather unconventional results regarding
these models are exposed in Secs. \ref{sec:resultsSM} and
\ref{sec:resultsXY}.

From a methodological point of view, we have provided a novel parallel
algorithm to Monte Carlo sample systems with $p=4$-body interactions
in an efficient way (i.e., in a time $\order(\nq)$ instead of
$\order(N\,\nq$)) in the unfavorable situation in which the
interaction network is non-sparse.

Moreover, we have stressed that these results, presented in
a statistical physical framework, have experimental consequences in
the field of photonics as these models describe also the interaction
between electromagnetic modes is passive mode locking lasers. In principle, they cover a broad range
of experimental circumstances in which the modes are subject to a
nonlinear quartic interaction and to a stochastic drift. In the case
of laser formation the drift is induced by the spontaneous emission,
considered as an effective thermal bath, and the different light
regimes are associated to different resulting thermodynamic phases of
the statistical model. This is well established in the case of the
mode locking transition of a closed cavity laser, which is solvable by
mean field theory in the so-called narrow-band
approximation.\cite{Gordon2002, Antenucci2014statistical} The present
work goes beyond mean field and allows to take into account frequency
correlations.  Our results not only account for general features of
discontinuous transitions observed in mode locking experiments, but
also predicts a variety of phenomenology as the vanishing of two-mode
correlations, the carrier phase delay of electromagnetic pulses or
the non-equipartite condensation, which presumably lies at the origin
of the experimental observations of Ref. \onlinecite{Leonetti2011},
and in this paper we determine the conditions under which these
phenomena arise. The model under analysis invites to establish further
links between the present results and other quantities measured in
laser experiments, as there are further quantities provided by the
Monte Carlo analysis that are experimentally accessible (intensity and
phase correlations and intensity spectra, hysteresis of the energy
curve). 

 This setup allows for an analysis with additional novel
ingredients as quenched interaction disorder,\cite{Conti2011} and any
type of interaction topology. This freedom is sufficient to enlarge
the spectrum of experimental situations that may be effectively
described in the statistical approach. It is a challenging problem
that of going beyond the passive mode locking transition in
establishing the link between optics and statistical physics. In other
words, in which circumstances a Hamiltonian formulation is possible
and what are the properties of the couplings 
appearing in Eq. (\ref{eq:Hgain}) describing a given experimental condition. 
Such a query is indeed a big theoretical challenge
which has motivated an intense research in the last
years, see Ref. [\onlinecite{AntenucciThesis}] for a review of the state of the art.
In random laser phenomena there is no
closed cavity and this fact poses several theoretical difficulties in
the treatment, as the very definition of lasing
mode;\cite{Dutra2000,Tureci2006} the presence of dissipative,
outer-radiative modes, and their effective influence in the set of
lasering modes;\cite{Viviescas2003} the possible existence of an
imaginary part in the coupling interaction; the existence of
correlations in the coupling disorder and, possibly, in the
noise.\cite{AntenucciThesis, Hackenbroich03} 

Besides the direct photonic interpretation, the Hamiltonian
Eq. (\ref{eq:H4new}) is quite general, and the form of topological
correlations (introduced as the FMC constraint, Eq. (\ref{eq:myFMC}))
is a very natural way of selecting the degrees of freedom which
effectively interact.  For this reason, we believe that the physical
consequences of the present study are not limited to optics, but are
possibly relevant in more general situations described by a scalar
field subject to a nonlinear interaction.



\section{Acknowledgments}

We thank Claudio Conti, Andrea Crisanti, Baruch Fischer, Neda Ghofraniha, Marco
Leonetti and Giorgio Parisi for motivating discussions.  The research leading to these
results has received funding from the Italian Ministry of Education,
University and Research under the Basic Research Investigation Fund
(FIRB/2008) program/CINECA grant code RBFR08M3P4 and under the
PRIN2010 program, grant code 2010HXAW77-008 and from the People
Programme (Marie Curie Actions) of the European Union's Seventh
Framework Programme FP7/2007-2013/ under REA grant agreement
n. 290038, NETADIS project.

\appendix

\section{The mean field solution of the ferromagnetic model}
\label{app:meanfield}

\noindent
Consider the fully connected ferromagnetic model
\begin{align}
\nonumber
& \mathcal{H} = 
 - \frac{1}{ N^3}   \sum_{j k l m}^{1,N}  a_{j} a_{k} a^\ast_{l} a^\ast_{m} \, ,
\\
& \text{with} \quad \sum_j |a_j|^2 = \epsilon N \, .
\label{spherical}
\end{align}
Defining $a_j = \sigma_j + i \tau_j$ the partition function is
\begin{align*}
 \mathcal{Z} =
 \int_{S}  \exp 
   \left(  
 \frac{\beta}{ N^3}  \sum_{jklm} \left( \sigma_{jklm}+\tau_{jklm} + \varphi_{jklm} \right)
 \right) \!
 d \boldsymbol \sigma 
 d \boldsymbol \tau
 \, ,
\end{align*}
where the subscript ${\cal S}$ means that the integral is evaluated
over the hyper-sphere Eq. (\ref{spherical}) and
\begin{align*}
 \varphi_{1234} &= \frac{1}{3} \left( \psi_{12,34}+\psi_{13,24}+\psi_{14,23} \right) \, , 
 \\
 \psi_{12,34} &= \sigma_{12} \tau_{34} + \sigma_{34} \tau_{12} \, ,
\end{align*}
and we are using the shortening 
\begin{align*}
  \sigma_{12 \ldots k} &= \sigma_1 \sigma_2 \cdots \sigma_k \, .
\end{align*}
Introducing the magnetizations 
\begin{align*}
 m_{\sigma} &= \frac{1}{N} \sum_j \sigma_j \, , &  m_{\tau} &= \frac{1}{N} \sum_j \tau_j \, ,
\end{align*}
the partition function is written as

%

\begin{align*}
  \mathcal{Z} &=
  \int \mathcal{D} \mathbf{m} \, 
 \,  e^{ -N  \beta  F(\mathbf{m})  } 
\end{align*}
with
\begin{align*}
 \beta F(\mathbf{m}) =
 - \beta  \left(
  m_\sigma^2 + m_\tau^2 
  \right)^2
  - \log \left[ \pi \left( \epsilon - m_\sigma^2 - m_\tau^2  \right) \right]
  - 1
 \, .
\end{align*}
Solving the integral over the magnetizations with the saddle point
method leads us to consider
\begin{align*}
 \beta \frac{d F}{d m_{\sigma,\tau}} = & \;
  2 m_{\sigma,\tau}  \left[
 - 2 \beta    \left( m_\sigma^2 + m_\tau^2 \right)
 +  \frac{1}{ \epsilon - m_\sigma^2 - m_\tau^2} 
 \right]
 = 0 \, .
\end{align*}
The paramagnetic (PM) case with $m_\sigma = m_\tau = 0$ is always a solution.
For 
\begin{align}
 \epsilon^2 \beta > 1
 \quad \to \quad 
 T < \epsilon^2 
\label{eq:FM_region}
\end{align}
also a ferromagnetic (FM) solution appears with
\begin{align*}
 m_\sigma^2 + m_\tau^2 =
 \frac{\epsilon}{2} \left( 1 + \sqrt{1-\frac{1}{\epsilon^2 \beta}} \right) \, .
\end{align*}
The average energy is 
\begin{align*}
 \langle \mathcal{H} \rangle  = 
 - \frac{\partial}{\partial \beta} \log \mathcal{Z}
 =  -  \epsilon^2  \left( m_\sigma^2 + m_\tau^2 \right)^2  + \mathcal{O}\left(\frac{1}{N}\right) \, ,
\end{align*}
and it is zero for the PM solution and
\begin{align}
  \frac{\langle \mathcal{H} \rangle}{N} & = - \frac{\epsilon^4}{4} \left(  1  + \sqrt{1 - \frac{1}{\epsilon^2 \beta}} \right)^2 \, ,
\end{align}
for the FM solution. 

The hessian of the functional $F$ yield the stability properties of the previous solutions.
The paramagnetic solution $m_\sigma=m_\tau=0$ is associated with two degenerate positive eigenvalues,
so the PM solution is always stable.
The FM solution has a \emph{null eigenvalue} and a positive eigenvalues,
then, in the region where the FM solution exists, it is always marginally stable.

Then, in the region of the phase diagram given by Eq. (\ref{eq:FM_region})
the stable PM and the marginal FM solutions coexist. 
The equilibrium transition is at the point
\begin{align*}
 \epsilon^2 \beta_c = 2.455408\ldots 
\end{align*}
where the free energy of the two solutions are equal. At lower
temperature $F_{\text{fm}} < F_{\text{pm}}$ and the PM solution
becomes metastable.


\section{Energy scaling in the disordered Spherical Model}
\label{app:energy}

We suppose a Gaussian distribution of couplings $P(J)$, with average
$J_0$ and variance $\sigma$. In this case, we can write an
$n$-replicated partition function:

\begin{eqnarray*}
  \overline{\mathcal{Z}^n} &=& \int \prod_{j=1}^N da_j~da_j^* 
    \int \prod_{[jklm]}J_{[jklm]}^{(4)} P(J_{[jklm]}) 
\\
&& \times\exp\left\{ -\beta
      \sum_{b=1}^n {\cal H}_J[\{a^{(b)}\}]\right\} 
=
 \\ 
   &=& \sqrt{2 \pi \sigma^2} \; \int  \prod_{j=1}^N da_j~da_j^* 
\\&&
\times \exp \Biggl\{
 \sum_{[jklm]}^{1,N}  \Biggl[ J_0 \beta\sum_{b=1}^n
      a_j^b a_k^{b*} a_l^{b} a_m^{b,*} 
\\
&&\qquad \quad + \frac{1}{2} \sigma^2 \beta^2
      \left(\sum_b a_j^{b} a_k^{b,*} a_l^{b} a_m^{b,*} \right)^2
      \Biggr]  ,
\end{eqnarray*}
where $[jklm]$ points out at distinct interacting quadruplets.  
Unlike the fully connected case, in a diluted case the ``spatial'' index of the
modes is not removed. However, just for scaling purposes, one can try and use a
mean-field approximation for the diluted case, as well, assuming that
$\sum_{N_4} \sim (\nq/N^4) \sum_{jklm}$, where the sum runs over all
indices. In this way one can rewrite the exponent in terms of the
overlap matrices and magnetizations as usual, so to obtain (cf. Ref. \onlinecite{Antenucci2014})
\begin{align*}
 \frac{E}{N} =& - \frac{1}{N} \frac{d}{d \beta} \overline{ \log \mathcal{Z}} = 
 - \frac{1}{N} \frac{d}{d \beta} \lim_{n \to 0} \frac{\overline{Z^n}-1}{n} =
 \\
=& -\frac{1}{2} \sum_b g(Q_{b1},R_{b1}) - k (m_\sigma,m_\tau) \, ,
\end{align*}
where ($a_1 \equiv \sigma_1 + i \tau_1$) 
\begin{align*}
Q_{ab} &= \sum_1 \frac{\sigma_1^a \sigma_1^b + \tau_1^a \tau_1^b }{2N} \, ,\\
R_{ab} &= \sum_1 \frac{\sigma_1^a \sigma_1^b - \tau_1^a \tau_1^b }{2N} \, ,
m_\sigma = \frac{1}{N} \sum_1 \sigma_1
\\
  g (Q_{ab},R_{ab}) &=
  \beta (Q_{ab}^2+R_{ab}^2) 
 \left[  \frac{1}{9}  \sigma_4^2 (Q_{ab}^2+R_{ab}^2) \frac{N_4}{N} \right] \, ,
 \\
 k(m_\sigma^a, m_\tau^a)& =   
 \frac{1}{2}  \left[  (m_\sigma^a)^2 +  (m_\tau^a)^2 \right] \\
& \left\{
 \frac{1}{12}  J_0^{(4)}  \left[  (m_\sigma^a)^2 +  (m_\tau^a)^2  \right] \frac{N_4}{N}
 \right\}
\, .
\end{align*}
In the case of equipartition, one has $\order(N)$ spins of
amplitude $\order(1)$, so all the overlap matrices and magnetizations
are $\order(1)$. Then the extensive energy in both cases results as in
Eq. (\ref{eq:Escalingdisorder}).

\bibliography{RLMbibliography}

\end{document}